\documentclass[a4report,11pt]{article}
\usepackage[dvips]{graphicx,rotating}
\usepackage{ae,lmodern}
\usepackage[utf8]{inputenc}
\usepackage[T1]{fontenc}
\usepackage{gensymb}
 \usepackage{epsfig}
    \usepackage{ulem}
 \usepackage[dvipsnames]{xcolor}
 \usepackage{amsmath}
 \definecolor{monRouge}{RGB}{255,0,60}
\newcommand{\patrick}[1]{\textcolor{red}{#1}}

\newcommand{\patrickf}[1]{\textcolor{red}{#1}}

 \definecolor{darkspringgreen}{rgb}{0.07,0.45,0.13}

\newcommand{\alphas}{\alpha_{s}}

\newcommand{\Dsm}{{D}_s^-}
\newcommand{\Dp}{{D}^+}

\newcommand{\Do}{{D}^0}

\newcommand{\Bb}{\overline{{B}}}

\newcommand{\pip}{\pi^{+}}
\newcommand{\pim}{\pi^{-}}

\newcommand{\Dstaro}{{D}^{\ast 0}}
\newcommand{\Dstarp}{D^{\ast +}}

\newcommand{\Vcb}{\left | {\rm V}_{cb} \right |}

\newcommand{\Vud}{\left | {\rm V}_{ud} \right |}
\newcommand{\Vus}{\left | {\rm V}_{us} \right |}

\newcommand{\Bs}{{B^0_s}}
\newcommand{\Bsb}{\overline{B}^0_s}
\newcommand{\Bp}{{B^{+}}}
\newcommand{\Bm}{{B^{-}}}

\newcommand{\Bob}{\overline{B}^0}
\newcommand{\Bd}{{B^{0}_{d}}}
\newcommand{\Bdb}{\overline{B}^{0}_{d}}

\newcommand{\Dsp}{D_s^+}

\newcommand{\Km}{{K}^-}
\newcommand{\Kp}{{K}^+}

\newcommand{\mumu}{\ifmmode {\mu^+\mu^-} \else ${\mu^+\mu^-} $ \fi}

\newcommand{\ba}{\begin{array}}
\newcommand{\ea}{\end{array}}
\newcommand{\bc}{\begin{center}}
\newcommand{\ec}{\end{center}}
\newcommand{\beq}{\begin{eqnarray}}
\newcommand{\eeq}{\end{eqnarray}}
\newcommand{\bes}{\begin{eqnarray*}}
\newcommand{\ees}{\end{eqnarray*}}
\newcommand{\Kz}{\ifmmode {\rm K^0_s} \else ${\rm K^0_s} $ \fi}
\newcommand{\Zz}{\ifmmode {\rm Z} \else ${\rm Z } $ \fi}
\newcommand{\qqbar}{\ifmmode {\rm q\bar{q}} \else ${\rm q\bar{q}} $ \fi}
\newcommand{\ccbar}{\ifmmode {\rm c\bar{c}} \else ${\rm c\bar{c}} $ \fi}
\newcommand{\bbbar}{\ifmmode {\rm b\bar{b}} \else ${\rm b\bar{b}} $ \fi}
\newcommand{\xxbar}{\ifmmode {\rm x\bar{x}} \else ${\rm x\bar{x}} $ \fi}
\newcommand{\rphi}{\ifmmode {\rm R\phi} \else ${\rm R\phi} $ \fi}

\begin{document}
\title{\bf \begin{huge} The $a_1$ factorisation coefficient in  $\Bob \to D^{(*)+}M^{-}$ and  $\Bob \to D^{(*)+}D_s^{(*)-}$ decays:   measurements versus theory. \ \\
\end{huge} }

\vspace{0.3cm}

\maketitle 


\begin{center}Alain Le Yaouanc\footnote{alain.le-yaouanc@ijclab.in2p3.fr}, Jean-Pierre Leroy\footnote{jean-pierre.leroy@ijclab.in2p3.fr}, Patrick Roudeau\footnote{patrick.roudeau@ijclab.in2p3.fr} \\
	Universit\'e Paris-Saclay, CNRS/IN2P3, IJCLab, 91405 Orsay, France


\end{center}

\begin{abstract}

Using recent measurements of exclusive B-meson decays we extract the $a_1$ factorization parameter in $\Bdb \to D^{(*)^+} K^-/\pi^-$ decay channels. { The  values obtained in the four channels are very similar, in agreement with the theoretical expectations obtained in the $m_Q \to \infty$ limit, but the measured  values differ definitely from the expected central values. } Such differences have already been observed. Using recent data, we improve the accuracy by {a factor close to  two} in this comparison. We study possible interpretations for such a difference and conclude that the claimed accuracy of corrections due to soft gluon contributions and finite mass effects has to be revisited before arguing for possible "New-Physics" effects. We observe that the corrections to the $m_Q \to \infty$ limit are larger in the spectator topology than  effects from exchange amplitude contributions.  We discuss also the way expected ratios of $a_1$ values, involving hypotheses from theory, are used to obtain the fraction of $\Bs$ meson production in jets at LHC and conclude that the uncertainties attached to this approach are not well established. Finally, the $a_1$ values extracted from  $\Bdb \to D^{(*)+} D_s^-$ decays are found to be similar, within uncertainties, to the ones obtained with $K^-$ emission, once expected penguin contributions are corrected. Using the same approach, we note that $a_1$ values measured in channels with a $D_s^{*-}$ emission are about two standard deviations lower than $a_1(D^{(*)+}K^-$).
\end{abstract}

\maketitle

\section{Introduction}

The factorization property of non-leptonic weak decays of heavy mesons has been, for over thirty years and under various forms, the object of extensive studies. On the theoretical side, an approximate expression for the amplitude of decays to heavy-light mesons is through the product of heavy-quark meson form factors and of the annihilation constant of the light meson. {The ratio of the exact result  to this approximation, is  expressed by means of the  parameter $a_1$,  the value of which would be $1$ in the absence of { further} strong interactions effects}. However a {reliable}  calculation of this parameter lies presently beyond our theoretical capacities, {except for the  {unphysical} $m_Q \to \infty$ limit}, and, consequently, one has to rely on phenomenological and experimental studies to get  a precise knowledge of its value.  

On the other hand, an accurate value for $a_1$ is, in turn, a necessary ingredient in the determination of $B_s$ decays through a quantity $f_s/f_d$ relating them to $B_d$ decays. It allows for example a precise determination of $B_s$ decay branching fractions at high energy colliders, and other quantities which are important to test the Standard Model (see below for more details). It is then essential that, {when doing so}, uncertainties (both from theory and experiment) be as small as possible and well under control. {Unfortunately, at present, there has appeared a definite and  large numerical discrepancy between the experimental results and  the theoretical expectations in the asymptotic limit, which are both very accurate on their own side}. It must be stressed that, while experimental studies are progressing much, theory is not able to provide an accurate evaluation of the $a_1$'s.
 
 In principle the deviation { from the calculable $m_Q\to \infty$ limit should be attributed to $1/m_Q$ corrections} but  no safe evaluation of these effects is presently available.
 

{ Since both
the theoretical expectations and the experimental data have evolved with time, we shall now { briefly review some} of the efforts that have  been devoted to these problems.}
\vspace{2mm}
 
 As to theory:
\begin{enumerate}
\item
in the seminal paper of Beneke, Buchalla, Neubert and Sachrajda (BBNS) \cite{ref:bbns}, the computations were performed at order NLO and the value of $a_1=1.06 \pm 0.02$ in the infinite quark mass limit was obtained. The authors have considered effects from $1/m_Q$ corrections and expected them to be small (a few percent). 

{Meanwhile, the latter computation of $1/m_Q$ corrections is rather crude and uncertain, and the authors were conscious of that.  Indeed, there is here a question of principle: {a quantitative calculation (that is using perturbative QCD and leaving aside lattice QCD which, here, is not tractable }) is only possible in the $m_Q \to \infty$ limit because, in this case,  
  only hard gluons  contribute to the exchanges between $B \to D$ and the emitted light meson {thanks to the occurrence of an exceptional  cancellation of the soft gluons. However, the latter remain in the   $1/m_Q$ corrections, a situation which impedes a safe calculation of the corrections, as underlined briefly after Eq. (28) of BBNS for the case of spectator quark corrections : the scale of   the strong coupling constant $\alpha_s$ in front of the correction is $k^2 \simeq \Lambda^2_{QCD}$, not a heavy scale, so that, as emphasised, the calculation cannot be safe.} 

 {After all, this should be no surprise, since the long demonstration of BBNS aims at showing that soft gluons can be eliminated in the asymptotic limit, opening the way to a systematic calculation, but by construction this holds only
in the limit. It was rightly considered  a remarkable success, precisely, that one could get rid of soft gluons but, obviously, these still keep raising a barrier away from this limit to a tractable quantitative treatment.}

{ Being quite aware of this barrier, the authors were formulating  rather prudent perspectives, relying mainly on future experimental progress rather than on theory. This is clearly seen for instance in the following quotation of the beginning of their section 4.}

{\it ``Rather, what we may do is ask whether data supports the prediction of a quasi-universal parameter |$a_1$| $\approx 1.05$ in these decays. If this is indeed the case, it would support the usefulness of the heavy-quark limit in analyzing non-leptonic decay amplitudes. If, on the other hand, we were to find large non-universal effects, this would point towards the existence of sizable power corrections to our predictions. We will see that with present experimental errors the data are in good agreement with our prediction of a quasi universal $a_1$ parameter. However, a reduction of the experimental uncertainties to the percent level would be very desirable for obtaining a more conclusive picture.''} 

But to understand these conclusions, one must recall that   at that time, around year 2000, the experimental value was $\left | a_1^{expt.}\right |=1.08 \pm 0.07$, which agreed with the asymptotic estimate but left room, also, for sizable $1/m_Q$ corrections.


\item
More recently, \cite{ref:huber_et_al} has extended the BBNS analysis in the infinite mass limit to a higher order in $\alphas$. A value slightly higher and more accurate than the previous one, $a_1 = 1.070 \pm 0.011$, was obtained. 

At the same time, the experimental values have decreased, becoming $\left | a_1^{expt.}\right |\simeq 0.90 \pm 0.05$ for the most accurate ones. Therefore there is a deviation at {the} $2-3\sigma$  {level} between experiment and the infinite mass limit evaluation \cite{ref:huber_et_al}. The authors {of this latter reference} have considered a possible contribution from $1/m_Q$ corrections, using the previous BBNS estimates. They say:

"\it{One possible interpretation of our findings that, on the one hand, the non-leptonic to semi-leptonic ratios come out larger in theory compared to experiment, and on the other hand the non-leptonic ratios in general agree with experiment, might be non negligible power corrections which could be negative in sign and 10 to 15 \% in size on the amplitude level. They would render the factorization test via non-leptonic to semi-leptonic ratios better, and at the same time could cancel out in the nonleptonic ratios, especially if they were of a certain universality." }}



\end{enumerate}

Experiments indicate that the corrections are quite large, in  contradiction with the expectations  of \cite{ref:facto1}.  However, as to us, we consider that   it is premature to resort to "New Physics",
and we rather consider, in view of all these discussions, that the crucial $1/m_Q$ corrections cannot be, at present, accurately estimated,  
and not even roughly,   {for lack of a systematic method}.

Note now that, at variance with $\left |a_1\right |$ absolute values, experimental results for ratios between $\left |a_1\right |$ values obtained in different reactions are close to 1, {as already observed}, suggesting a property of universality  of $a_1$.
This
is not unexpected, 
if applied to the so-called ``tree topology'' (T-topology). {
But it should be at least corrected by the contribution of the subleading ``$W$-exchange topologies'' to  part of the reactions.}

{  Unfortunately, as long as the knowledge
of the corrections remains only qualitative, it is difficult to attribute an uncertainty on those ratios, apart for those that are measured.}  {Therefore it is not appropriate to base the determination of $\Bs$ production at high energy colliders on techniques that are using universality of the $a_1$ parameter as we explain in Section \ref{discussionfs/fd}}

{
As announced above,
in
the case of light meson emission, $a_1$ universality
(up to some corrections like annihilation contribution)
is used to relate $\Bdb \to D^{(*)+}K^-/\pi^-$ and $\Bsb \to D_s^{(*)+}\pi^-$ decays in view of obtaining, at high energy colliders, a value for the ratio $f_s/f_d$ between the production fractions of strange and non-strange $\Bb$ mesons \cite{ref:lhcb_fsd} (see section \ref{discussionfs/fd} for a more detailed discussion). This then allows to evaluate absolute values of $\Bsb$ decay branching fractions by reference to $\Bdb$ decay branching fractions, measured at B-factories. In this exercise, we  consider that the corresponding uncertainties, expected from theory, are not really established.      In addition we find that the expression used to correct for the exchange in $\Bdb \to D^{(\star)+}\pim$ amplitudes in \cite{ref:fleisch1} is not correct.

{In the case of charm meson emission, the present study is related to our previous publication} \cite{ref:pepeetal} 
{on ${\cal B}(\Bb \to D^{**} \ell^- \bar{\nu}_{\ell})$ transitions,
in which we  study, {in addition},
$\bar{B} \to D^{**}D_s^{(*)-}$ decays.}

We consider more specifically Class I decays, i.e. $\Bdb \to D^{(*)+} M^-$ channels, in which the decay mechanism is dominated by the external emission, from the $b \to c$ quark transition, of a $W$ coupled to the meson $M$.
Quite surprisingly, the hypothesis works as well -or as roughly- for the emission of a charmed meson, contrarily to {the theoretical argument given in \cite{ref:bbns}.}



Our discussion will { address}  first 
the dominating external emission contribution. 
It seems to us that various definitions and statements require clarification{; an attempt to this is proposed in the following section.}

\subsection{Discussion of the factorization {hypothesis}}
The expression 
itself, ``factorisation hypothesis'', needs clarification. For Class I { processes}, it may be understood in a loose sense as meaning, for instance   in the case\footnote
{chosen here because it is the simplest case, being  purely Class I.} of $\Bdb \to \Dp \Km$, that the amplitude for the decay can be written as :
\begin{eqnarray}
{\cal M}= (G_F/\sqrt{2})~a_{1} \, ( m_B^2-m_D^2) \, f_{K}\, F_0(m_{K}^2) \label{eq:basic}
\end{eqnarray}
\noindent where $f_{K}$ describes the emission of the $K$ meson while $F_0(m_{K}^2)$ is the form factor of the $b \to c$ transition at $q^2=m_{K}^2$,
with a coefficient $a_{1}$ close to 1, which may depend on the particular transition.  ``Factorisation'' points to the fact that, approximately, the matrix element is given by  the product of simple hadronic factors. More precisely,  $a_1$  represents the effects of QCD which are not included in $f_K$ and $F_0$.   {Note that $a_{1}$ can be obtained by comparing measurements with expectations using Eq. (\ref{eq:basic})}. Meanwhile, in this case, the value of $a_1$ depends also directly on those of decay constants, form factors and $V_{cb}$.
 
 However,  there may be  additional contributions to the same process, the main one consisting in the exchange of a  $W$  between the two initial quarks. In this situation the initial $\Bb$ is annihilated, {in the sense that} there is no preserved (``spectator'') quark. This is the case,   for example,  when considering $\Bdb \to \Dp \pim$.    Then, traditionally, this contribution has been termed  ``annihilation'', but now one speaks mostly of ``exchange'' topology  and this is the expression we use in this note\footnote{ One must { stress that there exist} also pure exchange processes, such as  $\Bdb \to D_s^+ \Km$, which allow to scale this latter contribution.}. Finally there can also be  ``penguin'' and ``color-suppressed''\footnote{which however, for the latter, play no role in the present discussion.} contributions.  When considering such decays, we use the same expression for the decay amplitude as given in Eq. (\ref{eq:basic}), but we change the quantity $a_1$ into $a_{1,eff.}$. {The subscript  ``eff.'' refers to the above mentioned extra-contributions. Similar quantities, obtained from theory, will be indicated by using the superscript ``th.''.}

 Therefore we adopt the following notations for the various  { situations in which
the $a_1$ coefficient is considered:}
\begin{itemize}

\item   $a_{1}^{\infty,\, th.}(\Bb_Y,\,D_X M)$ is an estimate from theory, 
 in the $m_{Q}\to \infty$ limit. Consequently,  it does not include $W$-exchange  contributions, nor other $1/m_Q$ corrections (remaining inside the tree topology); 

\item  $a_{1,\, eff.}(\Bb_Y,\,D_X M)$ is the total measurable value in Class I processes where, possibly, additional decay amplitudes contribute as those mentioned above (such as { in} $\Bdb \to \Dp \pim$);

\item $a_{1}(\Bb,\,D_X M)$ is the measured value of the coefficient in processes described by the tree topology {alone} (as $\Bdb \to \Dp \Km$), in which case $a_{1}(\Bdb,\,D^+ K^-) \equiv a_{1,\, eff.}(\Bdb,\,D^+ K^-)$. 
{Also in those cases in which there are (subleading) contributions of other topologies,  it is the result one obtains by { subtracting} these contributions (as in $\Bdb \to \Dp \pim$ by subtracting the W-exchange contribution). This subtraction can be performed if the contribution can be evaluated separately as we show below for the exchange one, following the  approach of  \cite{ref:fleisch1}.}
\end{itemize}
 

The 
 statement that $\left |a_1\right |$ 
is ``close to 1'' is only qualitative. 
It has been the object of long efforts for finding a precise theoretical justification or even trying to calculate $a_1^{th.}$ --  this { attempt} culminating in BBNS \cite{ref:bbns}--  but one must warn from the beginning that in fact only a rough approximation is possible and that many illusions and misinterpretations have circulated. 
Even the minus sign of the departure from $1$ has not been predicted, {although it is likely that there is a negative contribution stemming from spectator $1/m_Q$ corrections \cite{ref:bbns}.} Empirical recipes
could  perhaps be formulated, but most probably they will also remain very approximate.

From the beginning, the hypothesis was supported by the idea that the soft gluons linking the $b \to c$ current to the pion or other emitted mesons could be neglected when the momentum transfer is large, whence all the non-perturbative physics would be included in the $F_0$ form factor and the pion annihilation constant.  But let us stress that, from the very idea, it was clear that an exact conclusion -if any- could be valid only in an asymptotic limit of large energy of the pion, and { can otherwise hold only as an approximation}, with unknown
{ uncertainty}.  

As stated previously, the most naive statement would be  $a_{1}^{th.}=1$. But, first, let us remark that hard gluons cannot be neglected anyway. {Actually} the basic weak interaction effective Hamiltonian reads :
\begin{eqnarray}
{\cal H}_W= (G_F/\sqrt{2})~(C_1 O_1+C_2 O_2) \label{eq:HW }
\end{eqnarray}
\noindent{where} $O_1=\bar{d}_{\alpha} \gamma_{\mu}(1-\gamma_5) u_{\alpha} ~~\bar{c}_{\beta} \gamma^{\mu}(1-\gamma_5) b_{\beta}$, $O_2=  \bar{d}_{\alpha} \gamma_{\mu}(1-\gamma_5) u_{\beta}~~ \bar{c}_{\beta} \gamma^{\mu}(1-\gamma_5) b_{\alpha}$
(with contracted greek indices $\alpha, \beta$ for colour) 
{and} the coefficients $C_1=1+{\cal O}(\alphas)$, $C_2={\cal O}(\alphas)$ containing effects of such hard gluons. Whence $a_{1}^{th.}=C_1+\frac{C_2}{N_c}=1+{\cal O}(\alphas)$. 
Before proceeding further, let us quote numbers including the leading correction at $\mu \simeq m_b$ \footnote{Those values are the ones obtained by  Buras \cite{ref:buras1994} in the  { t'Hooft-Veltman} scheme with $\Lambda_{\overline{MS}} = 225\, {\rm MeV} $ and $\alphas(M_Z) =  .117$.}:

$$ C_1 \simeq 1.1, ~C_2 \simeq -0.24,~ a_{1}^{th.} \simeq 1.03.$$


{ But it has been clear} {for long}
that this cannot be the full account of hard gluons. Indeed,
the above estimate of {$a_{1}^{th.}$} depends on the regularisation and renormalisation schemes, in particular, {it} would depend
 on the renormalisation point $\mu$. On the other hand, the physical amplitude ${\cal M}$ should  
{ of course show no such dependence}   (and the
factors $f_{K}, F_0(m_{\pi}^2)$ do not either). 

The correct expression for $a_{1}^{th.}$, in a well defined heavy quark mass asymptotic limit (made precise below), is systematically calculable  in perturbative QCD, up to distribution functions, which however give very small additional corrections, and a very accurate estimate has been given by BBNS {\cite{ref:bbns}}, including ${\cal O}(\alphas^2)$ contributions (further improved by Huber et al.,
 \cite{ref:huber_et_al}).
Numerically : $a^{\infty,\, th.}_{1}=1.070 \pm 0.011$ (with essentially no difference between an emitted pion or kaon).


Now, let us make explicit what 
{${\cal O}(1/m_Q)$ means}.  In the presentation of BBNS, it denotes {${\cal O}(1/m_{b,c})$} with the ratio $m_c/m_b$ kept fixed.  Another demonstration is given by Bauer, Pirjol and Stewart \cite{BPS} 
{ who require $m_b, m_c$ and  $E_{\pi}$ to be very large.} In any case, the limit is of course very far from the physical situation. In particular, it must be observed that in both cases, $m_c$ is required to be very large, which poses an obvious problem.
Moreover, as it is so often the case  in HQET, these  corrections 
{cannot be computed}, all the more for orders $(1/m_Q)^n, n>1$.
It must be stressed that{, at present,} the decay $\Bdb \to D^+ \Km$ can most probably not be treated in lattice QCD{, in view of the difficulty of this calculation}. Then, because of these ${\cal O}(1/m_Q)$ terms, the high precision : { {${\cal O}(\pm 0.01$),}} obtained in the asymptotic result  is actually of little help, since these terms cannot be evaluated systematically and cannot be expected to be very small. This high precision has even been {deceptive} since it has suggested that one could attain high precision in non leptonic decays. Rough estimates of 
${\cal O}(1/m_Q)$ corrections at lowest order, $n=1$, have been given by BBNS \cite{ref:bbns}, but anyway they meet objections as explained below\footnote{A first general problem is that, in order to estimate the corrections, they do not use the same expansion conditions as previously, namely they use a fixed, small value of $m_c$ while $m_b$ goes to $\infty$. This induces a change in the counting rules.}.

What is really ascertained 
is that the statement of factorisation hypothesis, i.e.  the proximity of $a_{1,\, eff.}$ to $1,$ has received a clear foundation. {On} the other side, it remains a qualitative statement \cite{ref:bbns}. This is clearly recognised by BBNS themselves, { as we said above; it may be that one is { indeed} too far from the asymptotic limit for it to be useful. 

\vskip 0.5cm

{\subsection{Exchange contributions}}

Among the ${\cal O}((1/m_Q)^n)$  {corrections}, beginning with $n=1$, there  are  the ``{exchange}'' ones. {This is in addition to the ``spectator contribution'' which remains within the tree topology.}
These 
contributions correspond to 
transitions in which  there is no ``spectator'' $\bar{d}$ quark but, instead,  the two initial $b \bar{d}$ quarks  annihilate 
{ against each other}. The general structure of the matrix element would then be  $<D \pi|J'_{\mu}|0><0|J'_{\mu}|\bar{B}>$,
where the $J'_{\mu}$ are either currents directly present in $O_{1,2}$ or ones generated by the Fierz transformation. 

Formally, the expression may seem similar to the factorisation, with the product of an annihilation constant and of a form factor. However, here, contrarily to the standard $m_Q=\infty$ factorisation contribution, the creation of a $\bar{d} d$ final pair to form the $D^+ \pi^-$ requires essentially a gluon emission by one of the $b,c,d,u$ quarks, whence at least an $\alphas$  factor.
Such a contribution is termed generically 
{ ``non-factorisable''}.


Attempts have been made to estimate the{ exchange contributions, which we shall discuss in confrontation with experiment, with the help of the pure exchange process $\Bdb \to \Dsp \Km$.} Obviously, these estimates are not at the same level of theory as the asymptotic result described above.

\vspace{2mm}



\subsection{Electro-weak corrections}

{
In the present analysis, our aim is to consider only, for clarity and simplicity, purely hadronic quantities, i.e. those deriving only from QCD, all the more since the question of electroweak corrections presents some complexity. On the other hand, we depend on experimental data, e.g. we use the measured non leptonic branching ratios, and then, for coherence, we should have to discard the electroweak effects from these data. But for reasons similar to the above mentioned ones, we do not attempt to proceed to this extraction and keep the experimental data as they are presented.} 

\patrick{}{Therefore, in $a_{1,\,eff.}$ evaluations, using nonleptonic and semileptonic $B$-decays we have not considered effects from radiative corrections that can be different for the two decay modes. Considering the present experimental accuracy attained in such branching fraction measurements, it seems of interest to evaluate, in the future, the real effects of these corrections, that can be of the percent order.}

\vspace{1cm}
\section{Factorization in light meson emission reactions}
\label{sec:factor_light}

First we {consider} the experimental evaluations of $a_{1,\, (eff.)}(\Bb_Y,\,D^{(*)+} M^-)$ in Class I nonleptonic decays in which the meson $M^-$ is light.
 
\subsection{Experimental determination of $a_{1,\, (eff.)}(\Bdb,\,D^{+} K^-/\pi^-)$}


In $\Bdb \to D^+ P^-$ channels, in which $P^-$ is a pseudo-scalar meson, the dominant decay mechanism corresponds to the external emission of a virtual W which is coupled to the emitted meson (tree process noted T in the following). In the factorization hypothesis the corresponding partial decay width is  
written as:
 
\begin{multline}
\Gamma(\Bdb \to \Dp P^-)= 6 \pi^2 C \left | a_{1,\,(eff.)}(\Bdb,\,D^+P^-) \right |^2\,|\vec{p}_D(m_P^2)| (f_P\,|V_{uq}|)^2 \\\times (m_B^2-m_D^2)^2  F_0^2(m_P^2)\label{eq:btodx_tree}
\end{multline}


\noindent with $C=G_F^2 \Vcb^2 /(96\,\pi^3\,m_B^2)$. $\vec{p}_{D}(m_P^2)$ is the $\Dp$ three-momentum evaluated in the $\Bdb$ rest frame, $f_P$  the charged meson leptonic decay constant and $V_{uq}$ the {appropriate} CKM matrix element{\footnote{If $B_d^0$'s are not separated from $\Bdb$ mesons,
contributions from $b \to u$ transitions are included in some of the measurements that are considered. 
In practice, {however, }this component can be neglected 
because it is doubly-Cabibbo suppressed {\it(DCS)}}.} 

{$F_0(q^2)$ -- which appears here for $q^2=m_P^2$ --} is a hadronic form factor which enters also in the $\Bdb \to D^+ \ell^- \bar{\nu}_{\ell}$ semileptonic decay rate through a term proportional to the charged lepton mass squared.
The hadronic current involves another hadronic form factor, $F_+(q^2)$,  which is measured in decays leading to a light lepton (see Eq. \ref{eq:dgdq2}):

\begin{equation}
\frac{d{\Gamma}}{dq^2}(\Bdb \to \Dp \ell^- \bar{\nu}_{\ell})= 4 C  m_B^2|\vec{p}_{D}(q^2)|^3 F_+^2(q^2)\label{eq:dgdq2}.
\end{equation}

Kinematics implies that $F_0(0)=F_+(0)$. For $q^2\neq 0,$ theoretical parameterizations of the $q^2$ dependence of hadronic form factors, whose parameters {have been fitted on data, are used \cite{ref:hfag2021}.}


{ The value of $  a_{1,\,(eff.)}(\Bdb,\,D^+P^-) $ is obtained by taking the ratio between Eq. (\ref{eq:btodx_tree}) and Eq. (\ref{eq:dgdq2})\cite{ref:Bjorken}:
 
\begin{multline}
\left | a_{1,\,(eff.)}(\Bdb,\,D^+P^-) \right |^2=\frac{{\cal B}(\Bdb \to \Dp P^-)}{\left.\frac{d{\cal B}}{dq^2}(\Bdb \to \Dp \ell^- \bar{\nu}_{\ell})\right|_{q^2=m_P^2}}\frac{1}{6 \pi^2}  \\ \times \frac{4\,m_B^2\,|\vec{p}_{D}(m_P^2)|^2 }{ (m_B^2-m_D^2)^2  (f_P\,|V_{uq}|)^2}\frac{F_+^2(m_P^2)}{F_0^2(m_P^2)}      \label{eq:a1dx_eff}.
\end{multline}
 
}
{ The result is independent {of} the value of $V_{cb}$. It does not depend either on the absolute values and parameterization of hadronic form factors because $F_+(m_P^2)\simeq F_0(m_P^2)$ when the light hadron is a $\pi$ or a $K$ meson. Moreover, the quantities $f_P\,|V_{uq}|$ are accurately measured for these hadrons.}

{ To evaluate $\vert a_{1,\,(eff.)}(\Bdb,\,D^+P^-)\vert$, in addition to the values and corresponding uncertainties (reminded below), one needs the parameters  which govern the $q^2$ variation of $F_+(q^2)$. We use  the ones  obtained with the CLN parameterization\footnote{{These results include charged and neutral semileptonic $\Bb$ decays measurements, which are combined by assuming that the corresponding semileptonic exclusive partial decay widths are equal.}}\cite{ref:hfag2021} :
\begin{eqnarray}
\eta_{EW}\,G_{1}(1)\,\Vcb & = & (41.53 \pm 0.98) \times 10^{-3}\nonumber\\
\rho^2 & = & 1.129 \pm 0.033 \nonumber\\
 C_{\eta_{EW}\,G_{1}(1)\,\Vcb,\,\rho^2} & = & 0.758\label{eq:btodlnumeas_HFLAV}
\end{eqnarray}}

\noindent{with ${\cal G}^2(w)= 4 r /(1+r)^2 F_+^2(w)$ and $r = m_D/m_B$; $w=(m_B^2+m_D^2-q^2)/2 m_B m_D$. For the 
form factor ratio,  we use the linear parameterization  $F_0(q^2)/F_+(q^2)=1-r_F\,q^2$, with $r_F= (0.022 \pm 0.001) GeV^{-2}$ as obtained in \cite{ref:bailey}. The input measurements are listed in Table \ref{tab:a1dk_input}. We have also rescaled the value obtained by integrating the CLN parameterization ($(2.13 \pm 0.07)\%$) such that it agrees with the value we are using for ${\cal B}(\Bdb \to \Dp \ell^- \bar{\nu}_{\ell})$ given in this table.}  

\begin{table}[!htb]
\begin{center}
  \begin{tabular}{|c|c|c|}
\hline
 &  PDG \cite{ref:pdg2022}& our estimate \\
\hline
${\cal B}(\Bdb \to \Dp \ell^- \bar{\nu}_{\ell})$  & $(2.25 \pm 0.08)\times 10^{-2}$ & $(2.18 \pm 0.06)\times 10^{-2}$\\
${\cal B}(\Bdb \to \Dp \pi^-)$ & $(2.56 \pm 0.08)\times 10^{-3}$ & $(2.57 \pm 0.08)\times 10^{-3}$\\
$R_{DK/D\pi}$ & $(8.19 \pm 0.20)\times 10^{-2}$ & $(8.19 \pm 0.20)\times 10^{-2}$\\
\hline
  \end{tabular}
  \caption[]{\it { Measurements used to evaluate $\left | a_{1,\,(eff.)}(\Bdb,\,D^+P^-) \right |$.
The value for ${\cal B}(\Bdb \to \Dp \ell^- \bar{\nu}_{\ell})$, in our estimate, is the average of charged and neutral $B$-meson decays, assuming that their semileptonic partial decay widths are equal. $R_{DK/D\pi}\equiv{\cal B}(\Bdb \to \Dp K^-)/{\cal B}(\Bdb \to \Dp \pi^-)$}
  \label{tab:a1dk_input}}
\end{center}
\end{table}

{In Table \ref{tab:a1dk_input}, the difference between the values quoted for 
${\cal B}(\Bdb \to \Dp \ell^- \bar{\nu}_{\ell})$ is explained by the fact that we are using the evaluation of this quantity obtained by averaging the measurements for  charged and neutral $\Bb$ mesons, assuming the equality of their corresponding partial decay widths. For ${\cal B}(\Bdb \to \Dp \pi^-)$, our result agrees with PDG2022 \cite{ref:pdg2022} and is more precise than the value quoted by HFLAV ($(2.56 \pm 0.13)\%$)\cite{ref:hfag2021}  which does not include a recent measurement from Belle \cite{ref:belle_dpi}.}

\subsubsection{$\Bdb \to \Dp \Km$}
\label{BdtoDpKm}

{ Because the decay $\Bdb \to \Dp \Km$ is described by the ``tree'' topology contribution alone (which does not mean, of course, tree approximation in the sense of quantum field theory since it includes gluon dressing), this process makes  it  possible, using Eq. (\ref{eq:a1dx_eff}), to measure directly $\left | a_{1}(\Bdb,\,D^+K^-) \right |$:}

{
\begin{equation}
\left | a_{1}(\Bdb,\,D^+K^-) \right |\equiv \left | a_{1,\,eff.}(\Bdb,\,D^+K^-) \right |= 0.877 \pm 0.023 \label{eq:a1dk}.
\end{equation}
{The quoted uncertainty is obtained by summing in quadrature the following
components: $\pm 0.018$ from {${\cal B}(\Bdb \to \Dp \pim)$  and $R_{DK/D\pi}$ measurements}, $\pm 0.014$ from fitted CLN parameters 
and $\pm 0.0015$ from $f_K \times V_{us}$.}} 
\vskip 0.5cm

What is striking in this measurement is the large magnitude of the deviation from the theoretical $m_Q \to \infty$ limit ({$\left |a_1^{\infty,\,th.}\right |-\left |a_{1}(\Bdb,\,D^+K^-)\right |=0.194 \pm 0.026$,  amounting to more than $7~\sigma$, given the high precision on both sides}).

This deviation is quite larger than the one announced by the theoretical estimates of BBNS {\cite{ref:bbns}}
for order ${\cal O}(1/m_Q)_{n=1}$ as given in their {Section $6.5$}. In fact for the $DK$ final state the only relevant contribution is the so-called spectator one. $T_{spec}/T_{lead}$ is given in their {Eq. (243)}, and amounts numerically to only $-3\%$ (the ratio should be roughly the same as for the $D\pi$ final state).

However, one must take into account the fact that the uncertainty affecting this theoretical result is probably itself very large. To support this, we will follow comments taken from the authors themselves. 

In the same {section}, {while} commenting {on} their numerical evaluations of the ${\cal O}(1/m_Q)_{n=1}$ corrections, they conclude : 
``{\it the typical size of power corrections to the heavy-quark limit is at the level of $10\%$ or less}''.  This means
 that they expect a large uncertainty, { all the more since,       in the $D^+\pim$ case which they are considering,} the spectator correction is counterbalanced by the exchange one, ending with a very small numerical estimate of the correction $-3+4=1 \%$, far below $10\%$.

A more explicit idea on the cause of such a large uncertainty in the evaluation of the ${\cal O}(1/m_Q)_{n=1}$
correction { to the spectator amplitude} is given below their  Eq. (28), 
 where they discuss their evaluation of the spectator correction. 
It is explained that their calculation, consisting in introducing a soft gluon through a propagator linking the hadron wave functions, cannot be ``justified'' since the momentum of the gluon (or ``virtuality'') is small, of order ${\cal O}(\Lambda_{QCD})$. Indeed, {it is clear that a perturbative treatment of soft gluons} is not possible. It   seems also reasonable to expect that the real effect of the soft gluon should be larger than expected from the naive insertion of the perturbative $\alphas$ at scale $m_b$. It is useful to note that the problem does not appear in the $m_Q \to \infty$ limit because, there, the soft gluons cancel against each other.

The authors of references \cite{ref:huber_et_al}  and \cite{ref:facto1}  
have updated the discussion of BBNS on the $1/m_Q$  corrections to the asymptotic limit, first by a more complete theoretical evaluation of this limit (higher orders in $\alphas$), secondly by new experimental estimates 
close to ours  (see Table \ref{tab:SiegenandUs}  in Section \ref{sec:heavymesons}), and finally,  in the second paper, by a more quantitative estimate of the matrix element in {Eq. (75)} of BBNS {\cite{ref:bbns}}, which controls the ``spectator'' $1/m_Q$ correction, and which they find to be very small. 
{Therefore, they conclude to a large
discrepancy between theory at finite mass and experiment, which would be of course puzzling. We do not try to discuss their calculation, but rather we just underline that LCSR results in general are very uncertain or affected with large errors even in the simplest cases (one may take as example the LCSR calculation of semileptonic $\Bb \to D^* \ell^- \overline{\nu}_{\ell}$ decay at $q^2=0$ by the authors of  \cite{ref:faller}). It must be much aggravated in the present case where the LCSR method is iterated, first at the level of a three particle distribution, then at the one of the calculation of the final matrix element for the $1/m_Q$
correction.  This is very clearly illustrated in the recent recalculation of the same quantity, with a similar method, by Piscopo and Rusov \cite{ref:piscopo} who give 
much more details : they conclude that 
error bands are very large, about $100 \%$ of the central value for the relevant matrix element $C_2 <O_2>$.
} 

\subsubsection{$\Bdb \to \Dp \pim$}

The decay $ \Bdb \to \Dp \pi^-$ is described by the sum of external $W$-emission and $W$-exchange graphs; therefore { what is measured is} $\left | a_{1,\,eff.}(\Bdb,\,\Dp\pim) \right |$ and, using {Eq. (\ref{eq:a1dx_eff}) and} the values listed in Table \ref{tab:a1dk_input}, one obtains:

\begin{equation}
\left | a_{1,\,eff.}(\Bdb,\,\Dp\pim) \right |=  0.846 \pm 0.019. \label{eq:a1dpieff}
\end{equation}
\noindent {The quoted uncertainty is obtained by summing in quadrature: $\pm 0.014$ from the $\Bdb \to \Dp \pim$ branching fraction measurement and $\pm 0.013$ from fitted CLN parameters. Other uncertainties can be neglected.}
\vskip 0.5cm
{The same observation can be addressed to this result as it has been above for $\Bdb \to \Dp \Km$ : the deviation is much larger than expected.}   {{In addition}, since the expected exchange contribution {is positive} ($+0.04$, {see Eq. (242)} of BBNS),  it is expected to nearly cancel the spectator one  {($-0.03$, see Eq. (243) of BBNS)}. {At variance { to this},  the measurements} indicate that the exchange contribution is much less important than the spectator correction. In Section \ref{sec:eval_exchange} we investigate in more details the exchange contribution.}

\subsection{Experimental determination of $a_{1,\, eff.}(\Bdb,\,\Dstarp \Km/\pim)$}
\label{sec:a1dstark}

We have repeated the same analyses as for $\bar{B} \to D$ transitions. {The values of $a_{1,\, eff.}(\Bdb,\,\Dstarp \Km/\pim)$ are obtained using ratios between corresponding nonleptonic and differential semileptonic branching fraction measurements, the latter being evaluated at $q^2=m_P^2$.}

Expressions relating helicity and Lorentz invariant form factors at $w=1$, which corresponds to $q^2 = q_{max}^2$, can be found in \cite{ref:fajfer}. We have used the CLN parameterization of these form factors
\cite{ref:cln} 
{ which} 
 depend on a single form factor, $h_{A_1}(w)$, with $w =(m_B^2+m_{D^{(*)}}^2-m_P^2)/(2\, m_B\, m_{D^{(*)}}) $, and on form factor ratios $R_{1,\,2,\,0}(w)$. Their variation versus $w$ is given in \cite{ref:cln} and \cite{ref:fajfer}.

{The fitted values of the different parameters, obtained by HFLAV \cite{ref:hfag2021}{, at w=1,}  are the following:}
{
\begin{eqnarray}
\eta_{EW}\,h_{A_1}(1)\,\Vcb & = & (35.00 \pm 0.36) \times 10^{-3}\nonumber\\
\rho^2 & = & 1.121 \pm 0.024 \nonumber\\
R_1(1) & = & 1.269\pm 0.026\nonumber \\
 R_2(1) & = & 0.853\pm 0.017\label{eq:btodstarlnumeas_HFLAV}.
\end{eqnarray}}

\noindent {The correlations between these quantities are given in \cite{ref:hfag2021}.}
The value for $R_0(1)$ and the corresponding uncertainty are obtained using the 
constraint  imposed by kinematics at $q^2=0$:
 
\begin{equation}
A_0(0) =  \frac{m_B+m_{D^*}}{2 m_{D^*}}A_1(0)-\frac{m_B-m_{D^*}}{2 m_{D^*}}A_2(0).
\label{eq:a3}
\end{equation}
{
\begin{multline}
\Gamma(\Bdb \to \Dstarp P^-)= 24\,\pi^2 C \left | a_{1,\,eff.}(\Bdb,\,\Dstarp P^-) \right |^2|\vec{p}_{D^*}(m_P^2)|^3\,\,m_B^2\,\\\times\left (f_P\,V_{uq}\,A_0(m_P^2) \right)^2.
\label{eq:btodstark}
\end{multline}
}
 

To evaluate $\vert a_{1,\,eff.}(\Bdb,\,\Dstarp P^-)\vert$ we use also, in addition to the values and corresponding uncertainties of the CLN parameters given in 
Eq. (\ref{eq:btodstarlnumeas_HFLAV}), the measurements given in Table \ref{tab:a1dstark_input}. { We rescale the 
{parameter fitted by HFLAV}, $\eta_{EW}\,h_{A_1}(1)\,\Vcb$, which fixes the normalisation for the decay rate  so that the integral of the CLN parameterization agrees with the value we have used for ${\cal B}(\Bdb \to \Dstarp \ell^- \bar{\nu}_{\ell})$.}

{
\begin{table}[!htb]
\begin{center}
  \begin{tabular}{|c|c|c|}
    \hline
 &  PDG \cite{ref:pdg2022}& our estimate \\
\hline
${\cal B}(\Bdb \to \Dstarp \ell^- \bar{\nu}_{\ell})$   & $(5.16 \pm 0.16)\times 10^{-2}$& $(5.17 \pm 0.13)\times 10^{-2}$\\
${\cal B}(\Bdb \to \Dstarp \pi^-)$ & $(2.74 \pm 0.13)\times 10^{-3}$ & $(2.64 \pm 0.11)\times 10^{-3}$\\
$R_{D^*K/D^*\pi}$ & $(7.74 \pm 0.36 )\times 10^{-2}$ & $(7.75 \pm 0.30)\times 10^{-2}$\\
\hline
  \end{tabular}
  \caption[]{\it { Measurements used to evaluate $\left | a_{1,\,(eff.)}(\Bdb,\,\Dstarp P^-) \right |$,
in addition to the parameters quoted in Eq. (\ref{eq:btodstarlnumeas_HFLAV}), relative to $\Bdb \to \Dstarp \ell^- \overline{\nu}_{\ell}$.  $R_{D^*K/D^*\pi}\equiv{\cal B}(\Bdb \to \Dstarp K^-)/{\cal B}(\Bdb \to \Dstarp \pi^-)$}. }
  \label{tab:a1dstark_input}
\end{center}
\end{table}
\noindent The difference observed on the values for ${\cal B}(\Bdb \to \Dstarp \pi^-)$ comes from the fact that 
{PDG \cite{ref:pdg2022} has not included a measurement from Belle \cite{ref:belle_dstarpi}; our result agrees with HFLAV \cite{ref:hfag2021}, $(2.63 \pm 0.10)\times 10^{-3}$, which has used this measurement.}}

\subsubsection{Results for $\Bdb \to \Dstarp \Km/\pim$}

Using the same procedure as for the $D^+ P^-$ channel and the  measurements listed in Table \ref{tab:a1dstark_input}, we obtain:

\begin{equation}
 \left | a_{1}(\Bdb,\,\Dstarp \Km) \right |\equiv \left | a_{1,\,eff.}(\Bdb,\,\Dstarp \Km) \right |= 0.870 \pm 0.027 \label{eq:a1dstark}
\end{equation}
\noindent{with a dominant contribution to the uncertainty ($\pm 0.025$) from the measurements of ${\cal B}(\Bdb \to \Dstarp \pim)$ and $R_{D^*K/D^*\pi}$. Other sources of uncertainties are $\pm 0.009$ from CLN parameters and $\pm 0.0015$ from $f_K \times V_{us}$.}

{In this channel also, the difference between the experimental result and the expected value is close to $20\%$ and is larger than $7 \sigma$.}

{For the decay $\Bdb \to \Dstarp \pim$, we obtain:}

\begin{equation}
 \left | a_{1,\,eff.}(\Bdb,\,\Dstarp \pim) \right |=  0.854 \pm 0.021 \label{eq:a1dstarpieff}
\end{equation}
\noindent{with a dominant contribution ($\pm 0.018$) to the uncertainty from the measurement of ${\cal B}(\Bdb \to \Dstarp \pim)$.
{{The comparison of the} measured values for  $ \left | a_{1,\,eff.}(\Bdb,\,\Dstarp \Km) \right |$ and $\left | a_{1,\,eff.}(\Bdb,\,\Dstarp \pim) \right |$ {shows clearly} that the difference} with the infinite quark mass prediction from theory comes mainly from spectator effects and not from the exchange contribution{, as observed already in  the corresponding decay channels with a $\Dp$}.  
}

\subsection{Evaluation of $W$-exchange 
amplitude contributions in $\Bdb \to D^{(*)+}\pi^-$ decays, $SU(3)$ assumptions.}
\label{sec:eval_exchange}
{We have shown, in the previous section, that the measured and expected values for $\left |a_1\right |$, in channels with an emitted kaon that are described by the T-topology alone, differ by large amounts and that this difference most probably indicates large ${\cal O}(1/m_Q)$ contributions, even if present evaluations of such corrections \cite{ref:facto1} appear to be very small, in contradiction with previous results \cite{ref:bbns} {that leave room for larger uncertainties on these evaluations}. In this section we compare the measured and expected results for the ratio between the exchange and the tree amplitudes.}

{The considered decays, with an emitted pion,} are the sum of external $W$-emission ($T$) and $W$-exchange $(E)$ amplitudes. {Therefore, using the fact that {the} corresponding decays with an emitted kaon are governed  by the sole $T$-topology, it is possible to evaluate the contribution from the $W$-exchange amplitude, once terms violating the SU(3) symmetry (such as $f_K$ used in place of $f_{\pi}$) are taken into account. {Moreover, one is helped by the existence of purely W-exchange processes, a circumstance which allows to fix directly the magnitude of this contribution.} In the following we have adopted the definitions and procedures described in \cite{ref:fleisch1}.  The results have been updated by using recent measurements, in particular for those that use the $\Bdb \to D^{(*)+}K^-$ decay.

{In addition, we obtain an expression for the correction factor to the pure T-topology  contribution which is the inverse of the one quoted in \cite{ref:fleisch1}.} 

{ We detail the evaluations for decays with a $\Dp$ while only results are quoted for $\Dstarp$ channels since the procedures are similar.}

\subsubsection{{ $\Bdb \to D^{+}\pi^-$}} 
\label{BDpiExch2}

{ Using the definitions for the $T$ and $E$ amplitudes given in \cite{ref:fleisch1},
we get:
\begin{multline}
\Gamma(\Bdb \to \Dp \pim) = 6 \pi^2 C |\vec{p}_D(m_{\pi}^2)|\,|V_{ud}|^2 (m_B^2-m_D^2)^2\\\times\left |T_{D\pi} + E_{D\pi} \right |^2.
\label{eq:btodpi_tpe}
\end{multline}
The expression for $T_{D\pi}$ is obtained by comparing the previous equation with
 {Eq. (\ref{eq:btodx_tree})}:
\begin{equation}
T_{D\pi} =  a_1(\Bdb,\,\Dp \pim)   f_{\pi} F_0(m_{\pi}^2).
\label{eq:btodpi_tdpi}
\end{equation}
The $a_1$ parameter is no {longer} said to be ``effective'' because it corresponds to the $T$-topology alone and, using Eq. (\ref{eq:btodx_tree}), this implies that:
\begin{equation}
\left | a_{1,\,eff.}(\Bdb,\,\Dp \pim) \right | =  \left | a_1(\Bdb,\,\Dp \pim) \right | \times  \left |1 + r_{D\pi} \right |
\label{eq:btodk_btodpi1}
\end{equation}
with $r_{D\pi}=\frac{E_{D\pi}}{T_{D\pi}}.$
}

The measured ratio $R_{D K /D\pi}$ is used to evaluate the contribution from the $W$-exchange amplitude in $\Bdb \to \Dp \pim$.

\begin{eqnarray}
R_{D K /D\pi}&\equiv\frac{{\cal B}(\Bdb \to \Dp \Km)}{{\cal B}(\Bdb \to \Dp \pim)}\hphantom{\left (\frac{f_K\, \Vus}{f_{\pi}\,\Vud} \frac{F_0(m_{K}^2)}{F_0(m_{\pi}^2)}\frac{ \left | a_1(\Bdb,\,\Dp \Km) \right | }{ \left | a_1(\Bdb,\,\Dp \pim) \right | }\frac{1}{\left | 1+ r_{D\pi}\right |}\right )}\nonumber\\
  \phantom{R_{D K /D\pi}}
& =\frac{|\vec{p}_{D}(m_K^2)|}{|\vec{p}_{D}(m_{\pi}^2)|}\left (\frac{f_K\, \Vus}{f_{\pi}\,\Vud} \frac{F_0(m_{K}^2)}{F_0(m_{\pi}^2)}\frac{ \left | a_1(\Bdb,\,\Dp \Km) \right | }{ \left | a_1(\Bdb,\,\Dp \pim) \right | }\frac{1}{\left | 1+ r_{D\pi}\right |}\right )^2.
\label{eq:btodk_btodpi}
\end{eqnarray} 

This gives:
{\it
\begin{equation}
\left | 1+ r_{D\pi}\right |=\frac{ \left | a_1(\Bdb,\,\Dp \Km) \right | }{ \left | a_1(\Bdb,\,\Dp \pim) \right | } \times (0.964 \pm 0.012).
\label{eq:esurt_dpi}
\end{equation}}

{The numerical quantity which appears in Eq.~(\ref{eq:esurt_dpi}) is simply the ratio between the values of $\left | a_{1,\, eff.}(\Bdb,\,\Dp \pim) \right | $
and $\left | a_1(\Bdb,\,\Dp \Km) \right | $, given in Eq. (\ref{eq:a1dpieff}) and Eq. (\ref{eq:a1dk}), respectively. {It is completely fixed by experiment. Its accuracy}
is better than the one on individual quantities  because the uncertainty on the semileptonic branching fraction does not contribute. { Moreover}, {the uncertainties stemming from the} form factors are also reduced because $F_0$ is evaluated at two nearby $q^2$ values and experiments have measured directly the ratio between the two decay modes.}

{To obtain more information on the ratio $|r_{D\pi}|=|E_{D\pi}|/|T_{D\pi}|$ and on the relative phase between the two amplitudes we follow the analysis {of }\cite{ref:fleisch1} and use the $\Bdb \to \Dsp \Km$ decay. This channel is a pure {$W$-exchange}  process but it differs from the {contribution entering}
$\Bdb \to \Dp \pim$ because an $s \bar{s}$ pair is created from vacuum in place 
of a $d \bar{d}$ pair. The corresponding partial decay width is expressed as \cite{ref:fleisch1}:}

 \begin{equation}
\Gamma(\Bdb \to \Dsp \Km)= \frac{G_F^2(m_B^2-m_{D_s}^2)^2}{16 \pi m_B^2}|\vec{p}_{D_s}(m_K^2)|\Vud^2 \Vcb^2\left |E_{D_s K} \right |^2.
\label{eq:btodsk}
\end{equation} 

In \cite{ref:fleisch1} the $W$-exchange amplitudes $E_{D_s K}$ and $E_{D\pi}$ are {linked} through the {relation}:
\begin{equation}
|E_{D_s K}| = |E_{D\pi}| \frac{ f_{K} f_{D_s}}{ f_{\pi} f_D}
\label{eq:ebtodpi_tdsk}
\end{equation}
{{which} can be derived from the BBNS expression given in  Section 6.5, Eq. (241)} of their paper \cite{ref:bbns}. {Again, it relies on} a definite way to treat $SU(3)$ breaking{, namely by assuming somewhat the validity of the expansion to order $(1/m_Q)^{n=1}$.}


{ The ratio between the branching fractions  of the $\Bdb$ into
$\Dsp \Km$ and $\Dp \pim$ becomes:} 
 
\begin{eqnarray}
R_{D_s K /D\pi}&\equiv& \frac{{\cal B}(\Bdb \to \Dsp \Km)}{{\cal B}(\Bdb \to \Dp \pim)\hfill}\nonumber\\
\phantom{R_{D_s K /D\pi} }& =& \frac{(m_B^2-m_{D_s}^2)^2}{(m_B^2-m_{D}^2)^2} \frac{|\vec{p}_{D_s}(m_K^2)|}{|\vec{p}_{D}(m_{\pi}^2)|}\left (\frac{f_K}{f_{\pi}}\frac{f_{D_s}}{f_D}\frac{\left | E_{D\pi}/T_{D\pi}\right |}{\left | 1+ E_{D\pi}/T_{D\pi}\right |}\right )^2.
\label{eq:btodsk_btodpi}
\end{eqnarray}

\noindent$B$-factories have measured the branching fraction ${\cal B}(\Bdb \to \Dsp \Km)$ whereas LHCb has measured directly the ratio, relative to the $\Dp \pim$
decay channel.
Because values obtained for $R_{D_sK/D\pi}$ at $B$-factories and LHCb  differ by about
{three} standard deviations the uncertainty quoted for the average of these two quantities is scaled according to the PDG recipe {\footnote{It consists in scaling the uncertainty on the average of N measurements by $\sqrt{\chi^2/(N-1)}$}}. This gives:
{
\begin{equation}
R_{D_sK/D\pi}=(1.12 \pm 0.22)\times 10^{-2}.\label{eq:btodsk_avg}
\end{equation}}

{To go further it is necessary to use additional hypotheses, from theory, because there are more unknowns ($r_{D\pi}$ modulus and phase, value of the ratio $r_{a_1} \equiv  \left | a_1(\Bdb,\,\Dp \Km) \right | / \left | a_1(\Bdb,\,\Dp \pim) \right |$ and possible corrections to the expression relating $\left |E_{D_s K} \right |$ and $\left |E_{D \pi} \right |$ in Eq. (\ref{eq:ebtodpi_tdsk}))  than the two constraints provided by experiment. In this respect, three situations are envisaged in the following paragraphs:}

\paragraph{analysis using the BBNS approach, including recent updates\\} 

{In the BBNS approach \cite{ref:bbns} 
{the} values for
$\left | a_1^{th.}(\Bdb,\,D^{(*)+}M^-)\right |$ are ``universal''. Thus one has:
\begin{equation}
\frac{\left | a_{1}(\Bdb,\,\Dp \pim) \right |} { \left | a_1(\Bdb,\,\Dp \Km) \right |} \simeq  \frac{\left | a_1^{th.}(\Bdb,\,\Dp \pim) \right |} { \left | a_{1}^{th.}(\Bdb,\,\Dp \Km) \right |} = 1.00  \pm (<0.01). \label{eq:a1_ratio_dpidk}
\end{equation}} 


{As explained before, Eq. (\ref{eq:ebtodpi_tdsk}) is also valid using this approach.} 

Therefore, using Eqs. (\ref{eq:esurt_dpi}), (\ref{eq:ebtodpi_tdsk}), (\ref{eq:a1_ratio_dpidk}) and the measured value for the ratio $R_{D_s K /D\pi}$ 
we obtain the modulus and the phase of the ratio $r_{D\pi}$:

\begin{equation}
\left |r_{D\pi} \right |=(7.3  \pm 0.7)\times 10^{-2},\,\phi(r_{D\pi}) =(121\pm 11)^{\circ}.
\label{eq:rdpi}
\end{equation}
{\noindent Because} the analysis is not sensitive to the sign of the imaginary part, we assume that it is positive.

The phase between the $E_{D\pi}$ and $T_{D\pi}$ amplitudes is higher than $90^{\circ}$ and close to this value.
The value for $\left |r_{D\pi} \right |$ is 
compatible with {the} initial estimates from \cite{ref:bbns} and {the} comments formulated in \cite{ref:huber_et_al}.

{{The phase, close to  $\pi/ 2$}, is more compelling : it completely contradicts
the typical feature of the BBNS approximation at this order, which gives real numbers. Let us emphasize that now, one goes beyond the known statement (Fleischer \cite{ref:fleisch1}) 
 that the phase does not vanish, as expected, in the $1/m_Q \to \infty$ limit. Apparently, predicting the phase  requires going beyond $1/m_Q$.}

{We have therefore considered {whether} it is possible to have a  phase equal to zero or $\pi$  between the two amplitudes by relaxing, in turn, each of the previously mentioned additional hypotheses.}

\paragraph{analysis without the constraint on the exchange amplitude\\}
Because ${\cal O}(1/m_Q)$ {terms} can be quite  important, they should  also enter  the expression which relates $E_{D_sK}$ to $E_{D \pi}$.
{In this test we still assume that {  the value of $a_1$, which originates from the T-topology contribution alone, is universal} and that $r_{D\pi}$ is real. 
{
However, we modify Eq. (20) by introducing the quantity $r_E$, that will correspond to such ${\cal O}(1/m_Q)$ corrections:
\begin{equation}
\left | E^{expt.}_{D_sK}\right | = r_E \, \left | E^{th.}_{D_sK}\right | = r_E \, \left | E_{D\pi}\right |  \frac{f_K f_{D_s}}{f_\pi f_D}
\end{equation}
and fit its value.}}

{We obtain:
\begin{eqnarray}
 r_{D\pi}=&(-3.6 \pm 1.2)\times 10^{-2}\\
 \,{\rm and\qquad} \,r_E=&2.0 \pm 0.7.
\end{eqnarray}
Thus, if Eq. (\ref{eq:ebtodpi_tdsk}) is modified by including a correction of ${\cal O}(1)$ given by the value of the parameter $r_E$, measurements can {accomodate} a real (negative) value for $r_{D\pi}$ at the few percent level.}


\paragraph{{analysis without assuming T-topology $a_1$ value universality\\}}
{We have redone the fit to data for the quantities, supposed to be real, $r_{D\pi}$ and 
$r_{a_1}= \left |a_1(\Bdb,DK)\right |/  \left |a_1(\Bdb,D\pi)\right |$, and using Eq. (\ref{eq:ebtodpi_tdsk}) in the constraint
provided by  $R_{D_s K /D\pi}$. We assume $r_{D\pi}$ to be negative; for positive values, the deviation from unity of $r_{a_1}$ would be larger.
We obtain:
\begin{eqnarray}
r_{D\pi}=&(-7.1 \pm 0.6)\times 10^{-2}\\
 \,{\rm and\qquad}\,r_{a_1}=& 1.037 \pm 0.015.
\end{eqnarray}
There is a $2 \sigma$ deviation from the universality.}

\subsubsection{$\Bdb \to D^{*+}\pi^-$}

{We proceed in full parallelism with the $\Dp \pim$ case (see the discussion in the previous section \ref{BDpiExch2}).}

The expression for $T_{D^*\pi}$ is:
\begin{equation}
T_{D^*\pi} = a_1(\Bdb,\,\Dstarp \pim)   f_{\pi} A_0(m_{\pi}^2)
\label{eq:btodstarpi_tdstarpi}
\end{equation}
and:

\begin{equation}
\left | a_{1,\,eff.}(\Bdb,\,\Dstarp \pim) \right | =  \left | a_1(\Bdb,\,\Dstarp \pim) \right | \times  \left |1 + r_{D^*\pi} \right |
\end{equation}
with $r_{D^*\pi}=\frac{E_{D^*\pi}}{T_{D^*\pi}}$.

\vspace{1mm}
The ratio $R_{D^*K /D^*\pi}$ is used to evaluate the contribution from the $W$-exchange amplitude in $\Bdb \to \Dstarp \pim$.  

\hspace{-2mm}
 
\begin{eqnarray}
R_{D^*K /D^*\pi} &=&\frac{{\cal B}(\Bdb \to \Dstarp \Km)}{{\cal B}(\Bdb \to \Dstarp \pim)}\\
&=&\left ( \frac{|\vec{p}_{D^*}(m_K^2)|}{|\vec{p}_{D^*}(m_{\pi}^2)|}\right )^3\left (\frac{f_K\, \Vus}{f_{\pi}\,\Vud} \frac{A_0(m_{K}^2)}{A_0(m_{\pi}^2)}\frac{\left | a_1(\Bdb,\,\Dstarp \Km) \right |}{\left | a_1(\Bdb,\,\Dstarp \pim) \right |}\frac{1}{\left | 1+ r_{D^*\pi}\right |}\right )^2\nonumber.
\label{eq:btodstark_btodstarpi}
\end{eqnarray}

This gives:
{
\begin{equation}
\left | 1+ r_{D^*\pi}\right |=\frac{ \left | a_1(\Bdb,\,\Dstarp \Km) \right | }{ \left | a_1(\Bdb,\,\Dstarp \pim) \right | } \times (0.987 \pm 0.020).
\end{equation}}


To have more information on the ratio $|E_{D^*\pi}|/|T_{D^*\pi}|$ and on the relative phase between the two amplitudes we follow the same approach as in section \ref{BDpiExch2}, using now the $\Bdb \to D_s^{*+} \Km$ decay. 

{ The $W$-exchange amplitudes, $E_{D_s^* K}$ and $E_{D^*\pi}$, are related through the new $SU(3)$ relation \cite{ref:fleisch1}, with the same justification as in the case of transitions to a pseudoscalar:}
\begin{equation}
|E_{D_s^* K}| = |E_{D^*\pi}| \frac{ f_{K} f_{D_s^*}}{ f_{\pi} f_{D^*}}.
\label{eq:ebtodstarpi_tdsstark}
\end{equation}

Therefore, the ratio between the branching fractions  of the $\Bdb$ into
$D_s^{*+} \Km$ and $\Dstarp \pim$ becomes: 

\begin{eqnarray}
R_{D_s^* K /D^*\pi} &=& \frac{{\cal B}(\Bdb \to D_s^{*+} \Km)}{{\cal B}(\Bdb \to \Dstarp \pim)}\nonumber\\
&=& \left (\frac{|\vec{p}_{D_s^*}(m_K^2)|}{|\vec{p}_{D^*}(m_{\pi}^2)|}\right )^3 \left (\frac{f_K}{f_{\pi}}\frac{f_{D_s^*}}{f_{D^*}}\frac{\left | r_{D^*\pi}\right |}{\left | 1+ r_{D^*\pi}\right |}\right )^2.
\label{eq:btodsstark_btodstarpi}
\end{eqnarray}
{\noindent For numerical evaluations, it has been assumed that $\frac{f_{D_s^*}}{f_{D^*}}=\frac{f_{D_s}}{f_{D}}$.}

$B$-factories have measured the branching fraction ${\cal B}(\Bdb \to D_s^{*+} \Km)$ and, dividing by the value of  ${\cal B}(\Bdb \to \Dstarp \pim)$ given in Table \ref{tab:a1dstark_input}, this gives:

{
\begin{equation}
R_{D_s^*K/D^*\pi}=(0.83 \pm 0.12)\times 10^{-2}.\label{eq:btodsstark_avg}
\end{equation}}



{Similarly to what we have done for pseudoscalar mesons,  we consider three possible  scenarios; the comments made above apply here too, {\it mutatis mutandis}.}

\paragraph{analysis using the BBNS approach, including recent updates\\}

{
\begin{equation}
\left |r_{D^*\pi} \right |=(6.7  \pm 0.5)\times 10^{-2},\,\phi(r_{D^*\pi}) =(107\pm 17)^{\circ}
\label{eq:rdstarpi}
\end{equation}}


\paragraph{analysis without the constraint on the exchange amplitude\\}

{
\begin{equation}
 r_{D^*\pi}=(-1.8 \pm 2.0)\times 10^{-2}\\,{\rm }\, r_{E^*}=3.7 \pm 4.2
\end{equation}
}

\paragraph{{analysis without assuming T-topology $a_1$ value universality\\}}
We obtain:
\begin{eqnarray}
r_{D^*\pi}=&(-6.4 \pm 0.4)\times 10^{-2} \\
{\rm and} \qquad r_{a_1^*}=&1.05 \pm 0.02.
\end{eqnarray}

\subsubsection{{Comparison with a previous similar analysis \cite{ref:fleisch1} and comments on $f_s/f_d$ measurement at LHCb}}
\label{discussionfs/fd}
The values we obtain for the quantity $\left | 1+ E_{D^{(*)}\pi}/T_{D^{(*)}\pi}\right |$
are compared with the results of \cite{ref:fleisch1} in Table \ref{tab:unplusesurt}. 
\begin{table}
\begin{center}
  \begin{tabular}{|c|c|c|}
    \hline
  & our analysis & \cite{ref:fleisch1}\\
\hline
$R_{D K /D\pi} \,(\%)$  & $8.19 \pm 0.20$ & $7.5 \pm 2.3$ \\
$\left | 1+ E_{D\pi}/T_{D\pi}\right |$  & $0.964 \pm 0.012$ & $1.01 \pm 0.15$\\
\hline
$R_{D^* K /D^*\pi} \,(\%)$  & $7.75 \pm 0.30$ & $7.76 \pm 0.45$ \\
$\left | 1+ E_{D^*\pi}/T_{D^*\pi}\right |$  & $0.987 \pm 0.020$ & $1.017 \pm 0.029$\\
\hline
  \end{tabular}
  \caption[]{\it { Comparison between the values, evaluated for the contribution of the {exchange} amplitude, in our analysis and in \cite{ref:fleisch1}. Values for the ratios $R$ are those from experiments, as available at the time of the two analyses.}
  \label{tab:unplusesurt}}
\end{center}
\end{table}

Once it has been assumed that all SU(3) corrections consist in changing the value of the light emitted meson decay constant, {the} values of $|1+E/T|$ depend only on the measured ratio of branching fractions $R_{D^{(*)} K /D^{(*)}\pi}$.
{ For decays where a  $\Dstarp$ is emitted, the uncertainty has been slightly reduced {thanks to the use of} new experimental data. It can be noted that, while values for $R_{D^{*} K /D^{*}\pi}$ are identical in the two analyses, the contribution from the exchange amplitude differs in sign. We wonder whether, in \cite{ref:fleisch1}, what was evaluated would not be $\left | 1+ E_{D^*\pi}/T_{D^*\pi}\right |^{-1}$ rather than the inverse.} 

{ For decays with an emitted $\Dp$, using measurements that were not available at the {time of \cite{ref:fleisch1}} { produces a reduction of the uncertainty} by an order of magnitude and the result becomes even more accurate than with an emitted $\Dstarp$. The contribution from the {$W$-exchange} amplitude is similar, for $\Dp$ and $\Dstarp$ channels, within uncertainties.} 

{The difference between  \cite{ref:fleisch1} and our analysis implies a significant effect on the correction factor used, at present, by LHCb to obtain the value for the ratio $f_s/f_d$ using the $\Bdb \to D^+\pim$ decay channel.} 

\noindent{\bf Comments on $f_s/f_d$ measurement at LHCb using {the $\Bdb \to D^{(*)+} \pim$} decay channel. } 
{To measure the $\Bsb$ relative to $\Bdb$ meson production rate at a hadron collider (usually noted as $f_s/f_d$), it has been proposed to use, among other methods, the corresponding decay channels $\Bsb \to \Dsp \pim$ and  $\Bdb \to \Dp \pim$ because, using factorisation, one can predict the ratio between these two branching fractions. Considering that the $\Bdb$ decay channel is not purely of the T-topology, it is necessary to correct for the effect of the exchange contribution. {For} this purpose the quantities ${\cal N}_{E^{(*)}}$ have been introduced \cite{ref:fleisch1}:
}
\begin{equation} 
{\cal N}_{E^{(*)}}= \left | \frac{1}{1+ E_{D^{(*)}\pi}/T_{D^{(*)}\pi}} \right |^2,
\label{eq:ne}
\end{equation}
\noindent for, respectively, the $\Bdb \to \Dp \pim$ and $\Bdb \to \Dstarp \pim$ {decay channels}.
The value, obtained in  \cite{ref:fleisch1}, from the analysis of the channel with an emitted $\Dstarp$ (Table \ref{tab:unplusesurt}) is ${\cal N}_{E^{*}}= 0.966 \pm 0.056$, from which they evaluate ${\cal N}_{E}= {\cal N}_{E^{*}}\pm 0.05$. The additional uncertainty is supposed to account for the fact that the correction, evaluated with an emitted $\Dstarp$, is used to correct the measurement of the decay $\Bdb \to \Dp \pim$,  with an emitted $\Dp$. In present LHCb analyses \cite{ref:fsoverfd_LHCb},  
{this procedure is still adopted, and they use ${\cal N}_{E}= 0.966 \pm 0.062$.} {The central value remains the same as in the previous evaluation and the total uncertainty too is similar.} 

{For the channel with an emitted $\Dstarp$, we obtain ${\cal N}_{E^{*}}= 1.027 \pm 0.042$.} 
{As already mentioned, the difference between the two  ${\cal N}_{E^{*}}$ central values   {most probably} originates from a confusion between ${\cal N}_{E^{*}}$ and $1/{\cal N}_{E^{*}}$ in  \cite{ref:fleisch1}.}


{In addition, it {can} be noted also that, thanks to progress in the experimental results, it is {no longer} needed to use decays with a $\Dstarp$ to correct measurements with an emitted $\Dp$. Using the corresponding values quoted in Table  \ref{tab:unplusesurt}, we obtain directly:  ${\cal N}_E= 1.076 \pm 0.027$, which, according to us, is the value that has to be used in practice (in place of $0.966 \pm 0.062$). It differs by $11\%$ from the one used at present and it is at least two times more accurate, with the benefit that there is no need to add an uncertainty (having an essentially arbitrary value) to cope for the difference between an emitted $\Dp$ or $\Dstarp$.}

{It is difficult to estimate the effect of such a correction on the present LHCb complex fitting procedure to obtain the value of $f_s/f_d$ \cite{ref:fsoverfd_LHCb} because there are also $\Bdb \to \Dp K^-$ (which has no annihilation contribution), semileptonic and $\Bp \to J/\Psi K^+$ decay channels that are used.}
 It can be noted that central values for $f_s/f_d$ are higher by about one standard deviation in fits with theoretical constraints \cite{ref:fsoverfd_LHCb}, indicating that {such} constraints, {currently used}, have some effect {thereon}.  

{According to the results given in Tables (9) and (10) of the LHCb publication \cite{ref:fsoverfd_LHCb}, the uncertainties on  $f_s/f_d$ are similar {whether one uses theoretical constraints or not}; therefore it seems more appropriate to use, at present, the values obtained without such constraints.} 

{The reason for this, in addition to the comments we made before about ${\cal N}_E$, is that in all these evaluations the equality 

$$\left | a_1(\Bdb,\,\Dp\pim) \right | =\left | a_1(\Bdb,\,\Dp\Km) \right | = \left | a_1(\Bsb,\,\Dsp\pim )\right |$$
{is assumed, although no well established uncertainty is available as long as there is no theoretical explanation for the large difference between the measured and the theoretical expected values for  $\left | a_1(\Bdb,\,\Dp\Km/\pim )\right |$.


{One {should} note also that, {although neither $\Bdb \to D^{(*)+} \Km$ nor $\Bsb \to D_s^{(*)+} \pim$ decays receive a} contribution from an exchange term, the spectator quarks are different for the two channels, being $d$ and $s$ respectively. The comparison between the values of  $\left | a_1(\Bdb,\,\Dp\pim) \right | $ and $\left | a_1(\Bdb,\,\Dp\Km) \right | $, which is sensitive to the effect of exchange terms, provides no information on the effect of spectator quarks with different flavours. Therefore, { the assumption} that $\left | a_1(\Bdb,\,\Dp\Km) \right | = \left | a_1(\Bsb,\,\Dsp\pim )\right |$ induces an uncertainty which is not controlled.}

\paragraph{Comments on  $f_s/f_d$ determination at LHCb using inclusive semileptonic decays.}

{From the discussion we have presented so far, one cannot {claim}, from the theory side, for precision physics in non-leptonic decays, as long as the origin of the discrepancy between the expected values and the measured ones for $\left |a_1\right |$ is not understood.}

{In this respect, using inclusive semileptonic decays, for which theory expects $\Gamma_{sl.}(\Bs)/\Gamma_{sl.}(\Bd)= (1-0.018 \pm 0.008)$ \cite{ref:gambino_bordone}, appears more promising. Meanwhile, in this case, the main difficulty comes from the experimental side because it is very difficult to measure such inclusive decay widths. Experiments select samples of events that are enriched, {respectively,} in strange and non-strange $\Bb$-meson decays and cross-contaminations are corrected. Most of these corrections are measured using the same apparatus but there are contributions from some channels that need to be corrected using external measurements or even guessed. {A typical example is} $\Bb \to \Dsp \Km X \ell^- \bar{\nu}_{\ell}$ decays which contaminate the  $\Dsp X \ell^-$, $\Bsb$ enriched sample. The corresponding correction is obtained using ${\cal B}(\Bm \to D_s^{(*)+}\Km \ell^-\bar{\nu}_{\ell} ) =(6.1 \pm 1.0)\times 10^{-4}$ measured at B-factories \cite{ref:pdg2022}. This is a quite small branching fraction but the correction to the selected $\Dsp X \ell^-$, amounting to $5-6\%$, is enhanced because strange $B$-mesons are much less abundant than non-strange ones. As can be noted, B-factories have measured only one exclusive decay channel with a $\Km$ produced in addition to the $D_s^{(*)+}$, {but there might be} additional pions. $b$-baryons can have similar decays with an hyperon produced in addition to the strange charm meson. It seems that such channels have not been considered in the present LHCb analysis. If one assumes, as a simple exercise, that  ${\cal B}(\Lambda_b \to D_s^{(*)+}Y \ell^-\bar{\nu}_{\ell} )= {\cal B}(\Bm \to D_s^{(*)+}\Km \ell^-\bar{\nu}_{\ell} )$ and that ${\cal B}(\Bm \to D_s^{(*)+} K \pi \ell^-\bar{\nu}_{\ell} )= 1/5 \times {\cal B}(\Bm \to D_s^{(*)+}\Km \ell^-\bar{\nu}_{\ell} )$,  the previous correction increases to $7-9\%$, depending on the $b$-hadron transverse momentum.}

\paragraph{ $f_s/f_d$ determination using exclusive semileptonic decays.}
{
The determination of the ratio $f_s/f_d$ depends on the knowledge, from theory, of the ratio between two decay branching fractions. As we have explained in the previous sections, uncontrolled uncertainties are attached to the ratios between exclusive nonleptonic channels as ${\cal B}(\Bdb \to \Dp \pim/\Km)/{\cal B}(\Bsb \to \Dsp \pim)$. The ratio between inclusive Cabibbo-allowed semileptonic decays of strange and non-strange $B$-mesons seems to be in better control from theory because the attached uncertainties, are estimated to be small as compared with the experimental ones.} 

{At present, and in particular because of recent progress in the evaluation of hadronic form factors in $\Bdb \to \Dp \ell^- \bar{\nu}_{\ell}$ and $\Bsb \to \Dsp \ell^- \bar{\nu}_{\ell}$ exclusive semileptonic decay channels {\cite{ref:facto1,ref:bordone_ffsl}}, it seems promising to measure the ratio $f_s/f_d$ using these transitions. As already demonstrated by LHCb \cite{ref:excl_Bdecays}, it is possible to use the decay channels $\Dp/\Dsp \to \Kp \Km \pip$ to obtain the same particles, in the final state, for the $\Bdb$ and $\Bsb$ mesons. In this way, a large amount of systematic uncertainties cancel in the ratio (trigger, particle identification, tracking efficiency, ...), when measuring $f_s/f_d$. The accuracy of this measurement will depend on the precision attained on charmed mesons decay branching fractions, which are, at present \cite{ref:pdg2022}: ${\cal B}(\Dp \to \Kp \Km \pip)= (9.68 \pm 0.18)\times 10^{-3}$ and ${\cal B}(\Dsp \to \Kp \Km \pip)= (5.38 \pm 0.10)\times 10^{-2}$. They correspond to a relative uncertainty of 2.6$\%$.} 

{Unfortunately, up to now, LHCb has used their measurements to obtain a value for
the CKM matrix element $V_{cb}$, in $\Bsb \to D_s^{(*)+} \mu^-\bar{\nu}_{\mu}$ decays, under the assumption that their determination of $f_s/f_d$ is well under control. In our proposal, it seems more appropriate to invoke the fact  that the value of $V_{cb}$ is universal and does not enter when considering the ratio between $\Bsb$ and $\Bdb$ semileptonic decays, to obtain a value for $f_s/f_d$.}
 
{\begin{equation}
\frac{f_s}{f_d}= \frac{N^{rec.}_{B_s\to D_s\mu\nu}}{N^{rec.}_{B_d\to D\mu\nu}}\frac{\int{d\Gamma(B_d\to D\mu\nu)/dw \, \epsilon_{B_d\to D\mu\nu}(w) dw}}{\int{d\Gamma(B_s\to D_s\mu\nu)/dw \, \epsilon_{B_s\to D_s\mu\nu}(w) dw}}\frac{\tau(B_d)}{\tau(B_s)}\frac{{\cal B}(D^+)}{{\cal B}(D_s^+)}
\end{equation}}
 
In this expression, the first ratio corresponds to the measured $\Bsb$ and $\Bdb$ exclusive semileptonic decays. The second term is the ratio between the integrated semileptonic rates, taken from theory, mutiplied by the corresponding reconstruction efficiencies. At the end of this expression, there is the ratio between the $\Dp$ and $\Dsp$ branching fractions, into the $\Kp \Km \pip$ decay channel. It is not possible for us to evaluate the part, in this expression, that depends on the experiment performances. The same expression can be used, replacing $D$ by $D^*$ mesons and considering more kinematic variables in the integrals.
}

\section{$\left |a_1(\Bsb, D_s^+ \pi^-)\right |$ absolute determination}
{Accurate evaluations of the $\left |a_1\right |$ parameters in several $\Bb$ decay channels may help to understand the large differences with corresponding theoretical determinations.}

{A possibility to evaluate $\left |a_1(\Bsb, D_s^+ \pi^-)\right |$ is to use the same approach as in Section \ref{sec:factor_light}.}

{For this purpose, we use recent results from LHCb \cite{ref:fsoverfd_LHCb}:} 
{ 
\begin{eqnarray}
{\cal B}(\overline{B}^0_s \to D_s^+ \pi^-)\times 10^{3}&=&3.20 \pm 0.10 \pm 0.16 \,\,\,(3.02 \pm 0.10)\label{eq:bsds}\\
{\cal B}(\overline{B}^0_s \to D_s^+ \mu^- \overline{\nu}_{\mu})\times 10^{2}&=&2.40 \pm 0.12 \pm 0.15 \pm0.06 \pm 0.10 \, (2.24 \pm 0.18).\nonumber 
\end{eqnarray} 
{These values are corrected to remove the common uncertainty coming from the evaluation of $f_s/f_d$. The central values and the corresponding uncertainties are slightly modified when they depend on the values of external decay branching fractions, to be in agreement with PDG or with the corresponding values used in previous sections. The value of ${\cal B}(\overline{B}^0_s \to D_s^+ \pi^-)$ depends on ${\cal B}(\overline{B}^0_d \to \Dp \pi^-)$ for which we use the value given in Table \ref{tab:a1dk_input}. The value of the exclusive semileptonic $\Bsb$ decay branching fraction depends on the values of ${\cal B}(\overline{B}^0_d \to \Dp \mu^- \overline{\nu}_{\mu})$ (Table \ref{tab:a1dk_input}) and of the $\Dp/D_s^+$ decay branching fractions to $\Kp\Km\pip$ \cite{ref:pdg2022}.} The values obtained in this way are quoted within parentheses in Eq. (\ref{eq:bsds}).}
{ To extract from the experimental data the  value of}  $\left |a_1(\Bsb, D_s^+ \pi^-)\right |$, one needs also the values of the CLN parameters, fitted on the differential distribution
$d{\cal B}(\overline{B}^0_s \to D_s^+ \mu^- \overline{\nu}_{\mu})/dq^2$, which have been published in \cite{ref:excl_Bdecays}. We have
integrated the corresponding expression and scaled the result such that it agrees with the quoted measured branching fraction in Eq. (\ref{eq:bsds}).}

{Using an expression similar to the one quoted in Eq. (\ref{eq:a1dx_eff}), which was relative to the decay of the $\Bd$ meson
we obtain:
\begin{equation}
\left | a_{1}(\Bsb,\,\Dsp \pim)\right |= 0.922 \pm 0.026.
\end{equation}
This result does not depend on the value assumed for the ratio $f_s/f_d$. The main contributions to the uncertainty are those which stem from the semileptonic and nonleptonic branching fraction measurements: $\pm 0.021$ and $\pm 0.015$, respectively. 
The value of  $\left | a_{1}(\Bsb,\,\Dsp \pim)\right |$ can be compared with the one obtained previously for $\left | a_{1}(\Bdb,\,\Dp \Km)\right |= 0.877 \pm 0.023$, which is expected to be the same according to \cite{ref:fleisch1}. Present results agree with this expectation: $\left | a_{1}(\Bsb,\,\Dsp \pim)\right |-\left | a_{1}(\Bdb,\,\Dp \Km)\right |= 0.045 \pm 0.034$.}}

\section{Factorization in heavy meson emission reactions}\label{sec:heavymesons}


{According to \cite{ref:bbns}, non-leptonic Class I decays of the type {$ \Bb \to D^{(*)} D_s^{(*)-}$ in which a charmed meson is emitted in place of a light one are not expected to satisfy factorization, { i.e. one does not expect $\left |a_1(\Bb, D^{(*)} D_s^{(*)-})\right |$ to be (even roughly) close to $1$, because one does not expect $\left |a^{\infty,th}_1(\Bb, D^{(*)} D_s^{(*)-})\right |=1$ { in the $m_Q \to \infty$ limit.}} Moreover, in these decays, penguin amplitudes are contributing and their effect has to be estimated before comparing with factorization estimates.
}

{Conversely to decays in which a light meson is emitted, the effects of the non-dominant amplitudes are not expected to give a significant difference between charged and neutral $B$-mesons decays because:
\begin{itemize}
\item the penguin amplitudes, denoted  "$P$" in the following, are the same;
\item in final states with a $D_s^{(*)-}$, the {exchange}  amplitude, which contributes in charged $B$ decays only, is DCS.
\end{itemize}    
{Therefore, the 
measurements with a charged or a neutral $B$-meson can be averaged (and, in the following, we express the result in terms of the neutral state).} 
We have evaluated our own averages using existing measurements, updating values of intermediate decay branching fractions and taking into account correlations. 
}

{Within the factorization framework, one can use a  formulation similar to the one of Section \ref{sec:eval_exchange} to express the effect of penguin contributions by defining:
\begin{multline}
\left | a_{1,\,eff.}(\Bdb,\,D^{(*)} D_s^{(*)-}) \right |\\ =  \left | a_1(\Bdb,\,D^{(*)} D_s^{(*)-}) \right | \times  \left |1 + r_{D^{(*)} D_s^{(*)-}} \right |
\label{eq:btodds_btodk1}
\end{multline}
\noindent with $r_{D^{(*)} D_s^{(*)-}}=\frac{P_{D^{(*)} D_s^{(*)-}}}{T_{D^{(*)} D_s^{(*)-}}}$. Values for the {induced} correction to the  $\left |a_1\right |$ parameter, $\left |1 + r_{D^{(*)} D_s^{(*)-}} \right |$, have been evaluated in \cite{ref:cheng_yang} and are reminded {of} in Table \ref{tab:a1_cheng}:
 \begin{center}{
\begin{table}[!htb]
\begin{center}
  \begin{tabular}{|c|c|c|}
    \hline
 decay channel & $\left |1 + r_{D^{(*)} D_s^{(*)-}} \right |$ & $\left | a_{1,\,eff.}(\Bdb,\,D^{(*)} D_s^{(*)-}) \right | / \left | a_{1}(\Bdb,\,D^{(*)}\Km) \right |$\\
    &  \cite{ref:cheng_yang} & as obtained in our analysis\\            
\hline
$\Bdb \to \Dp D_s^-$  & $0.847$ & $0.901 \pm 0.054$ \\
$\Bdb \to \Dstarp D_s^-$  & $1.037$ & $1.155 \pm 0.075$ \\
$\Bdb \to \Dp D_s^{*-}$  & $0.962$ & $0.850 \pm 0.068$ \\
$\Bdb \to \Dstarp D_s^{*-}$  & $0.962$ & $0.838 \pm 0.042$\\ 
\hline
  \end{tabular}
  \caption[]{\it { Expected corrections to the $\left |a_1\right |$ parameter in channels with an emitted $D_s^{(*)-}$ {as} obtained in \cite{ref:cheng_yang}. In the last column, similar quantities are given, under the hypothesis that $\left | a_{1}(\Bdb,\,D^{(*)} D_s^{(*)-}) \right | = \left | a_{1}(\Bdb,\,D^{(*)}\Km) \right |$ (see the following sections).}
  \label{tab:a1_cheng}}
\end{center}
\end{table}
}\end{center} 

In the following, for {each one of } the different decay channels that are considered, we repeatedly use the following procedure:
\begin{itemize}
\item{we first estimate, from the experimental data,  the ratio $\frac{\left | a_{1,eff.}(\Bdb,\,D^{(*)} D_s^{(*)-}) \right |}{\left | a_{1}(\Bdb,\,D^{(*)}\Km) \right |}$ where the $K^-$-channel, in the denominator,  has been chosen because it involves no exchange contribution. The result is then compared with the estimated effects of the penguin amplitudes as obtained in   \cite{ref:cheng_yang} 
assuming  that $\left | a_{1}(\Bdb,\,D^{(*)} D_s^{(*)-}) \right | = \left | a_{1}(\Bdb,\,D^{(*)}\Km) \right |$, which implies that factorization applies also to decays with heavy meson emission;}
\item then we provide values for  $\left | a_{1,\, eff.}(\Bdb,\,D^{(*)} D_s^{(*)-}) \right |$ using, in addition to the nonleptonic decay channels, the {corresponding} semileptonic ones, as was already done  in Section \ref{sec:factor_light} in the case of light meson emission.
\end{itemize} 

{We use the expressions for ${\cal B}(\overline{B} \to D^{(*)} D_s^{*-})$ derived in \cite{ref:pepeetal}.}
}

\subsection{The $\overline{B} \to D D_s^-$ decay channel}
\label{sec:a1ddbar}

The branching fractions used in this analysis are given in Table \ref{tab:bddbar_avg}.
\patrickf{
\begin{table}
\begin{center}
  \begin{tabular}{|c|c|c|c|}
    \hline
 decay channel& PDG \cite{ref:pdg2022}& HFLAV \cite{ref:hfag2021}& our average\\
\hline
${\cal B}(\Bdb \to \Dp D_s^-)\times 10^{3}$  & $7.24 \pm 0.77$ & $7.82 \pm 0.72$ & $7.30 \pm 0.72$\\
${\cal B}(\Bm \to \Do D_s^-)\times 10^{3}$  & $9.01 \pm 0.94$ & $9.58 \pm 0.98$ & $10.0 \pm 1.35$\\
\hline
  \end{tabular}
  \caption[]{\it { Branching fractions for $\Bb$ meson decays into a $D_s^-$ and another charm pseudo-scalar meson, obtained in PDG 2022, HFLAV 2022 and in our analysis. }
  \label{tab:bddbar_avg}}
\end{center}
\end{table}}
The smaller uncertainty obtained for $\Bdb$ than for $\Bm$ {results from the fact that}, for $\Bdb$-mesons only, there is an additional result from Belle {\cite{ref:belle-DDs}} which is more accurate than other measurements. { Our uncertainty for the $\Bm$ is also higher because we have not included the ratio between the results obtained by LHCb for neutral and charged $B$-meson decays}, having considered that it should correspond to the one expected from the corresponding lifetimes, from which it differs by less than two sigmas. Therefore this is simply a 
{ consistency} check of the LHCb analysis. 


From the values given in Table \ref{tab:bddbar_avg} and assuming that their partial decay widths are equal, we combine results for charged and neutral $B$-mesons to obtain:

\begin{equation}
{\cal B}(\Bdb \to \Dp D_s^-)= (7.76 \pm 0.84)\times 10^{-3}
\label{eq:br_B_to_DDs} 
\end{equation}

\noindent{in }which the quoted uncertainty has been scaled by 1.37, according to the PDG recipe.

{
\subsubsection{Penguin contribution (step 1  of the procedure explained above)}
To evaluate the penguin amplitude contribution we compare the  decay channel under consideration with $\Bdb \to \Dp \Km$.
{ In oher words,} we consider
the ratio $\vert a_{1,\,eff.}(\Bdb,\,\Dp D_s^-) / a_{1}(\Bdb,\,\Dp \Km)\vert$ obtained by using 
an expression similar to Eq. (\ref{eq:btodk_btodpi}):}

\begin{eqnarray}
\frac{ \left | a_{1,\,eff.}(\Bdb,\,\Dp D_s^-) \right | }{ \left | a_1(\Bdb,\,\Dp \Km) \right | }
&=&\sqrt{\frac{{\cal B}(\Bdb \to \Dp D_s^-)}{{\cal B}(\Bdb \to \Dp K^-)}\frac{|\vec{p}_{D}(m_K^2)|}{|\vec{p}_{D}(m_{D_s}^2)|}}\,\frac{f_K\, \Vus}{f_{D_s}\,|V_{cs}|} \,\frac{F_0(m_{K}^2)}{F_0(m_{D_s}^2)}\nonumber\\
&=& 0.901 \pm 0.054.
\label{eq:btoddbar_btodk}
\end{eqnarray}

\noindent{This value agrees within one standard deviation with the expectation from theory : $0.847$ \cite{ref:cheng_yang}. }


\subsubsection{$\left | a_{1,\,eff.}(\Bdb,\,\Dp \Dsm) \right | $ evaluation (step 2)}

{ We evaluate $\left | a_{1,\,eff.}(\Bdb,\,\Dp D_s^-) \right |$ using the semileptonic decay channel as in Eq. (\ref{eq:a1dx_eff}):
 \begin{multline}
\left | a_{1,\,eff.}(\Bdb,\,\Dp \Dsm) \right |^2=\frac{{\cal B}(\Bdb \to \Dp \Dsm)}{\left.\frac{d{\cal B}}{dq^2}(\Bdb \to \Dp \ell^- \bar{\nu}_{\ell})\right|_{q^2=m_{D_s}^2}}\frac{1}{6 \pi^2}\\\times\frac{4\,m_B^2\,|\vec{p}_{D}(q^2)|^2}{ (m_B^2-m_D^2)^2  (f_{D_s}\,|V_{cs}|)^2 }\frac{ F_+^2(m_{D_s}^2)}{      F_0^2(m_{D_s}^2)}\label{eq:a1dds_eff}
\end{multline} 

\noindent {which gives:
\begin{equation}
\left | a_{1,\,eff.}(\Bdb,\,\Dp \Dsm) \right |= 0.790 \pm 0.046\label{eq:a1dds_eff2}
\end{equation}
in which the dominant {contributions to the uncertainty} are : $\pm 0.043$ from the ${\cal B}(\Bdb \to \Dp \Dsm)$ measurement, $\pm 0.009$ from the $f_{D_s}\,|V_{cs}|$ determination and  $\pm 0.013$ from the CLN parameterization of semileptonic form factors.}


 After extraction of the penguin contribution, expected from theory:
\begin{equation}
\left | a_{1}(\Bdb,\,\Dp D_s^-) \right |= (0.790\pm 0.046)/0.847=0.933\pm 0.054.
\label{eq:br_B_to_DDs} 
\end{equation}}

{The result is compatible, within one standard deviation,  with the number obtained from light meson emission:  $\left | a_{1}(\Bdb,\,\Dp K^-) \right | = 0.877 \pm 0.023$.} 




\subsection{The $\overline{B} \to D D_s^{*-}$ decay channel}
\label{sec:dv}


{ We consider now decay channels with the emission of a vector meson instead of a pseudo-scalar one and averaging, as for the previous channel,  measurements obtained with charged and neutral $B$-mesons we obtain :}



{
\begin{table}[!htb]
\begin{center}
  \begin{tabular}{|c|c|c|c|}
    \hline
 decay channel& PDG \cite{ref:pdg2022}& HFLAV \cite{ref:hfag2021} & our average\\
\hline
${\cal B}(\Bdb \to \Dp D_s^{*-})\times 10^{3}$  & $7.4 \pm 1.6$ & $7.5 \pm 1.9$ & $ 7.8\pm 1.6$\\
${\cal B}(\Bm \to \Do D_s^{*-})\times 10^{3}$  & $7.6 \pm 1.6$ & $8.3 \pm 2.0$ & $ 7.8\pm 1.5$\\
\hline
  \end{tabular}
  \caption[]{\it { Branching fractions for $\Bb$ mesons decaying into a $D$ and a strange charm vector meson, obtained in PDG 2022, HFLAV 2022 and in our analysis. }
  \label{tab:bddstarbar_avg}}
\end{center}
\end{table}
}

{
{\it
\begin{equation}
{\cal B}(\Bdb \to \Dp D_s^{*-})= (7.45 \pm 1.07)\times 10^{-3}.
\label{eq:br_B_to_DDsstar} 
\end{equation}
}}


{ 
\subsubsection{Penguin contribution}
To evaluate the penguin amplitude contribution we have compared, similarly, this decay channel with $\Bdb \to \Dp \Km$. Therefore we consider
the ratio $\vert a_{1,\,eff.}(\Bdb,\,\Dp D_s^{*-}) / a_{1}(\Bdb,\,\Dp \Km)\vert$ obtained by using Eq. (\ref{eq:btodv}) to compute the partial decay width for the decay $\Bdb\to \Dp D_s^{*-}$.}

 \begin{multline}
\Gamma_{Tree}(\Bdb \to \Dp V^-)
           = 24 \pi^2 C | a_{1}(\Bdb,\,\Dp V^-) |^2\,|\vec{p}_D(m_V^2)|^3\, (f_V\,|V_{q_1q_2}|)^2  \\ \times  \,m_B^2\, F_+^2(m_V^2)
\label{eq:btodv}
\end{multline} 

\noindent where $q_1$ and $q_2$ are a generic  notation for the appropriate quarks constituting the  specific vector meson under consideration.
This gives:


{ 
\begin{eqnarray}
\frac{ \left | a_{1,\,eff.}(\Bdb,\,\Dp V^-) \right | }{ \left | a_1({\Bdb,\,\Dp \Km} \right | })
&= \sqrt{\frac{{\cal B}(\Bdb \to \Dp V^-)}{{\cal B}(\Bdb \to \Dp K^-)}\frac{|\vec{p}_{D}(m_K^2)|\,(m_B^2-m_D^2)^2}{4\,|\vec{p}_{D}(m_{V}^2)|^3\,m_B^2}}\,\frac{f_K\, \Vus}{f_{V}\,|V_{q_1q_2}|}\nonumber  \\&\times  \frac{F_0(m_{K}^2)}{F_+(m_{V}^2)} \\   
&={ 0.850 \pm 0.068.   \label{eq:a1dv_a1dk}}
\end{eqnarray}
}

\noindent For the $D_s^*$ decay constant we use $f_{D_s^*}/f_{D_s}=1.26 \pm 0.03$ \cite{ref:damir_fdstar}. 


{The dominant contributions to the uncertainty on the obtained ratio are  $\pm 0.063$ from the branching fraction measurement, $\pm 0.022$ from  the decay constants and CKM elements determination.
The value obtained in Eq. (\ref{eq:a1dv_a1dk}) is about two standard deviations lower than the expectation from theory : $0.962$ \cite{ref:cheng_yang}.
}
\subsubsection{$\left | a_{1,\,eff.}(\Bdb,\,\Dp D_s^{*-}) \right | $ evaluation}

 We evaluate $\left | a_{1,\,eff.}(\Bdb,\,\Dp D_s^{*-}) \right |$ using the semileptonic decay channel as in Eq. (\ref{eq:a1dx_eff}):
\begin{align}\begin{split}
\left | a_{1,\,eff.}(\Bdb,\,\Dp D_s^{*-}) \right |^2=\frac{{\cal B}(\Bdb \to \Dp D_s^{*-})}{\left.\frac{d{\cal B}}{dq^2}(\Bdb \to \Dp \ell^- \bar{\nu}_{\ell})\right|_{q^2=m_{D_s^{*}}^2}}
 \frac{1}{6 \pi^2}\frac{1}{ (f_{D_s^*}\,|V_{cs}|)^2}\label{eq:a1ddsstar_eff}
\end{split}\end{align}


{\noindent {which gives:
\begin{equation}
\left | a_{1,\,eff.}(\Bdb,\,\Dp D_s^{*-}) \right |= 0.746 \pm 0.058\label{eq:a1ddsstar_eff2}
\end{equation}
in which the dominant {contribution to the} uncertainty is  $\pm 0.055$  and comes from the ${\cal B}(\Bdb \to \Dp D_s^{*-})$ measurement.}}


{{ After extraction of the penguin contribution expected from theory one obtains:
\begin{equation}
\left | a_{1}(\Bdb,\,\Dp D_s^{*-}) \right |= (0.746\pm 0.058)/0.962=0.775\pm 0.060.
\label{eq:br_B_to_DDsstar} 
\end{equation}}}

{ The result is about two standard deviations lower than the light meson emission value:  $\left | a_{1}(\Bdb,\,\Dp K^-) \right | = 0.877 \pm 0.023$.}


\subsection{The $\Bb \to D^{*+} \Dsm$ decay channel}
\label{sec:a1dstardbar}
We have repeated the same analysis as in Section \ref{sec:a1ddbar} for $\Bb \to D^*$ transitions. {The measured branching fractions are listed in Table \ref{tab:bdstardbar_avg}}.

{
\begin{table}[!htb]
\begin{center}
  \begin{tabular}{|c|c|c|c|}
    \hline
 decay channel &PDG \cite{ref:pdg2022}& HFLAV \cite{ref:hfag2021} & our average \\
\hline
${\cal B}(\Bdb \to \Dstarp D_s^-)\times 10^{3}$  & $8.0 \pm 1.1$ & $8.1 \pm 1.4$ & $9.1 \pm 1.2$\\
${\cal B}(\Bm \to \Dstaro D_s^-)\times 10^{3}$  & $8.2 \pm 1.7$ & $7.9 \pm 1.6$ & $8.1 \pm 1.7$\\
\hline
  \end{tabular}
  \caption[]{\it { Branching fractions for $\Bb \to D^* D_s^-$  decays obtained in PDG 2022, HFLAV 2022 and in our analysis. }
  \label{tab:bdstardbar_avg}}
\end{center}
\end{table}
}
{ Our value for the neutral $B$-meson differs from the other averages because of the way we correct the measurements from BaBar.}
From the values given in Table \ref{tab:bdstardbar_avg} and assuming that 
{their} partial decay widths are equal, we combine results for charged and neutral $B$-meson to evaluate:

{
\begin{equation}
{\cal B}(\Bdb \to \Dstarp D_s^-)= (8.6 \pm 1.0)\times 10^{-3}.
\label{eq:br_B_to_DDs} 
\end{equation}
}
\subsubsection{Penguin contribution}
Using Eq.(\ref{eq:btodstark_btodstarpi}) {we evaluate the ratio} :


{
\begin{align}\begin{split}
\frac{ \left | a_{1,\,eff.}(\Bdb,\,\Dstarp D_s^-) \right | }{ \left | a_1(\Bdb,\,\Dstarp \Km) \right | }
&=\sqrt{\frac{{\cal B}(\Bdb \to \Dstarp D_s^-)}{{\cal B}(\Bdb \to \Dstarp K^-)}\frac{|\vec{p}_{D^*}(m_K^2)|^3}{|\vec{p}_{D^*}(m_{D_s}^2)|^3}}\,\\& \times\frac{f_K\, \Vus}{f_{{D_s}}\,|V_{cs}|}  \frac{A_0(m_{K}^2)}{A_0(m_{D_s}^2)}\\
&= 1.155 \pm  0.075.
\label{eq:btodstardbar_btodstark}
\end{split}\end{align}}




The value obtained  agrees with the theoretical estimate -$1.037$ \cite{ref:cheng_yang}- which includes contributions from the penguin amplitude in $a_{1,\,eff.}(\Bdb,\,\Dstarp D_s^-)$.



{
\subsubsection{$\left | a_{1,\,eff.} (\Bdb, \,\Dstarp \Dsm )\right |$ evaluation}}
{We evaluate $\left | a_{1,\,eff.}(\Bdb,\,\Dstarp \Dsm) \right |$ using the semileptonic decay channel, as in Section \ref{sec:a1dstark}:

\begin{equation}
\left | a_{1,\,eff.}(\Bdb,\,\Dstarp \Dsm) \right |= 1.000 \pm 0.066 \label{eq:a1dstards_eff2}
\end{equation}
\noindent {in which the dominant uncertainties are  $\pm 0.059$ from the ${\cal B}(\Bdb \to \Dstarp \Dsm)$ measurement, $\pm 0.011$ from the  $f_{D_s}\,|V_{cs}|$ determination and  $\pm 0.027$ from the CLN parameterization of semileptonic form factors.}}

{\noindent After extraction of the penguin contribution 
one obtains:
\begin{equation}
\left | a_{1}(\Bdb,\,\Dstarp D_s^-) \right |= (1.000\pm 0.066)/1.037=0.964\pm 0.064.
\label{eq:br_B_to_DstarDs} 
\end{equation}
The result is not significantly different from the light meson emission value~: $0.870 \pm 0.027$ for $\Dstarp \Km$.}

\subsection{The $\Bdb \to \Dstarp D_s^{*-}$ decay channel}
\label{sec:dstarv}
If, in place of a pseudo-scalar, we consider decay channels with vector meson emission, the corresponding decay width becomes:
{
\begin{align}\begin{split}
\Gamma_{Tree}(\Bdb \to \Dstarp V^-)= &6 \pi^2 C  \left | a_{1}(D^*V) \right |^2\, (f_V\,|V_{q_1q_2}|)^2 \\\times m_V^2\,m_{D^*}&\,m_B|\vec{p}_{D^*}(m_V^2)|  FF_{D^*V}^2(m_V^2)
\label{eq:btodstarv}
\end{split}\end{align}

\noindent with: 

\begin{align}\begin{split}
FF_{D^*V}^2(m_V^2)  =  A_1^2(m_V^2)&\frac{(1+r)^2}{r} \left (3+\frac{m_B^2|\vec{p}_{D^*}(m_V^2)|^2}{ m^2_V m^2_{D^*}} \right ) +  \\
 \frac{4 r}{(1+r)^2}   \frac{|\vec{p}_{D^*}(m_V^2)|^2}{m_{D^*}^2}   &\left(V^2(m_V^2)   + A_2^2(m_V^2)   \frac{m_B^2|\vec{p}_{D^*}(m_V^2)|^2}{2 m^2_V m^2_{D^*}}\right) \\
 -   4 A_1(m_V^2)&\, A_2(m_V^2)\frac{m_B^2|\vec{p}_{D^*}(m_V^2)|^3}{ m^2_V m^3_{D^*}}
\label{eq:ffdstarv}
\end{split}\end{align}

\noindent in which $r=m_{D^*}/m_B$.

{The measured branching fractions for the corresponding decay channels are listed in Table \ref{tab:bdstardstarbar_avg}.}


\patrickf{
\begin{table}
\begin{center}
  \begin{tabular}{|c|c|c|c|}
    \hline
 decay channel& PDG \cite{ref:pdg2022}& HFLAV \cite{ref:hfag2021}  & our average \\
\hline
${\cal B}(\Bdb \to \Dstarp D_s^{*-})\times 10^{3}$  & $17.7 \pm 1.4$ & $18.7\pm1.5$ & $ 17.8\pm 1.3$\\
${\cal B}(\Bm \to \Dstaro D_s^{*-})\times 10^{3}$  & $17.1 \pm 2.4$ & $ 17.9 \pm 2.7$ & $17.6 \pm 2.4$\\
\hline
  \end{tabular}
  \caption[]{\it { Branching fractions for $\Bb$ mesons decaying into a $D^*$ and a strange charm vector meson, obtained in PDG 2022, HFLAV 2022 and in our analysis. }
  \label{tab:bdstardstarbar_avg}}
\end{center}
\end{table}
}
Assuming that their partial decay widths are equal, we combine results for charged and neutral $B$-mesons to evaluate:}
{
\begin{equation}
{\cal B}(\Bdb \to \Dstarp D_s^{*-})= (17.5 \pm 1.1)\times 10^{-3}.
\label{eq:br_B_to_DDs} 
\end{equation}}

\subsubsection{Penguin contribution}
{It is evaluated by considering the ratio of $\left | a_{1,\,eff.} (\Bdb, \,\Dstarp D_s^{*-} )\right |$ and $\left |a_1(\Bdb, \,\Dstarp K^-)\right |$:
{
\begin{eqnarray}
\frac{ \left | a_{1,\,eff.}(\Bdb,\,\Dstarp V^-) \right | }{ \left | a_1(\Bdb,\,\Dstarp K^-) \right | }
 &=&2\,\sqrt{\frac{{\cal B}(\Bdb \to \Dstarp V^-)}{{\cal B}(\Bdb \to \Dstarp K^-)}\frac{m_B\,|\vec{p}_{D^*}(m_K^2)|^3}{|\vec{p}_{D^*}(m_{V}^2)|\,m_V^2\,m_{D^*}}}\nonumber\\
&\times&\frac{f_K\, \Vus}{f_{V}\,|V_{q_1q_2}|} \,\frac{A_0(m_{K}^2)}{FF_{D^*V}(m_{V}^2)}\nonumber\\
&=& 0.838 \pm 0.042.
\label{eq:a1dstarv_a1dstark}
\end{eqnarray}}}





{
\subsubsection{$\left | a_{1,\,eff.}( \Bdb, \,\Dstarp D_s^{*-} )\right |$ evaluation}}

{We evaluate $\left | a_{1,\,eff.}((\Bdb,\,\Dstarp D_s^{*-}) )\right |$ using the semileptonic decay channel, as in Section \ref{sec:a1dstark}:
\begin{equation}
\left | a_{1,\,eff.}(\Bdb,\,\Dstarp D_s^{*-}) \right |= 0.729 \pm 0.033 \label{eq:a1dstardsstar_eff2}
\end{equation}
\noindent in which the dominant {sources of uncertainty} are : $\pm 0.023$ from the ${\cal B}(\Bdb \to \Dstarp D_s^{*-})$ measurement, $\pm 0.015$ from the CLN parameterization of semileptonic form factors, $\pm 0.008$ from $f_{D_s}\,|V_{cs}|$ determination, and $\pm 0.017$ from the $f_{D_s^*}/f_{D_s}$ estimate.}

{\noindent After extraction of the penguin contribution, expected from theory, one obtains:
\begin{equation}
\left | a_{1}(\Bdb,\,\Dstarp D_s^{*-}) \right |= (0.729\pm 0.033)/0.962= 0.758 \pm 0.034.
\label{eq:br_B_to_DstarDsstar} 
\end{equation}
This value differs by more than three standard deviations from  : $\left | a_{1}(\Bdb,\,\Dstarp K^-) \right |= 0.870\pm 0.027$.}

\section{Comparison between measurements and theoretical expectations}

\begin{table}[!htb]
\begin{center}
  \begin{tabular}{|l|c|c|}         
\hline
   &Huber[2]&this work\\
   \hline
$\vert a_{1, eff.}(\Bdb, D^{+}\pi^-)\vert$  & $0.89 \pm 0.05$ & $0.846 \pm 0.019$ \\
$\vert a_{1, eff.}(\Bdb, D^{*+}\pi^-)\vert$  & $0.96 \pm 0.03$ &$0.854\pm 0.021$ \\
$\vert a_{1, eff}.(\Bdb, D^+K)\vert$  & $0.87 \pm 0.06$ & $0.877 \pm 0.023$ \\ 
$\vert a_{1, eff}.(\Bdb, D^{*+}K)\vert$  & $0.97 \pm 0.04$ & $0.870 \pm 0.027$ \\ 
\hline
  \end{tabular}
  \caption[]{\it {  Comparison between the different determinations of the $\vert a_1\vert$ parameter obtained from experimental measurements in [2] (2016) and in our analysis (2022).}}
  \label{tab:SiegenandUs}
\end{center}
\end{table}

\subsection{Values for $| a_1 |$ with an emitted light meson}

{The statement of factorisation as a qualitative one -i.e. that $a_1$ is not {very} far from $1$- is reasonable, but it does not  have the status of a definite quantitative estimation, and theory is unable to provide it.} 

The values we obtain from experimental data, 
$\vert a_{1, eff.}(\Bdb, D^+K^-)|=0.877 \pm 0.023$ and $\vert a_{1, eff.}(\Bdb, D^{*+}K^-)|=0.870 \pm 0.027$} are, on the other hand,  significantly lower than the theoretical expectation in the infinite $m_Q$ limit: $a_1^{\infty,th.}(\Bdb,D^+K^-)=1.055 \pm 0.018$ at NLO or $a_1^{\infty,th.}(\Bdb,D^{*+}K^-)=1.070 \pm 0.011$ at NNLO. The difference amounts to seven standard deviations and therefore is very well established. 

There is no problem in observing a difference, since of course we are not in the theoretical limit : there should be $1/m_Q$ corrections. 

The real problem is twofold : 

1) As in most processes of $B$ physics except the simplest ones, there does not exist presently a tractable, safe and systematic theoretical method to evaluate exact amplitudes from QCD.
The only one in principle is lattice QCD, but it
is far outside our present capacities for these NL decays.

2) Various evaluations of the $1/m_Q$ corrections to the known $m_Q \to \infty$ limit by approximate methods have been proposed, but they suffer from serious doubts. In fact they disagree between themselves and they are far from the observed values.




{ The $1/m_Q$ corrections have been considered by BBNS in the framework of semi-perturbative QCD methods involving distribution functions \cite{ref:bbns} and the authors believed them to be small.  There would remain however to
include ${\cal O}(1/m_Q^n)$  corrections (with $n>1$). Moreover, as stressed by the same authors  themselves, all these are controlled by soft gluon effects, which are beyond a rigourous treatment. 
{The attempt of \cite{ref:facto1}, making iterated use of LCSR, concludes to a very small $1/m_Q$ spectator correction, contradicting strongly experimental data, but we think that the conclusion cannot  be trusted because of the very large uncertainties which must  be expected
in this method (see the indications of
the end of Section \ref{BdtoDpKm} for a further discussion, and in particular the reference to the more recent paper of Piscopo and Rusov \cite{ref:piscopo}).}


Therefore we consider that there is presently no compelling reason to invoke
 New Physics effects to explain such differences as is done in \cite{ref:facto1}; present uncertainties on  ${\cal O}(1/m_Q^n)$ corrections leave room for a standard explanation.

 Such differences and discrepancies with observation must not cast doubt on the whole theory. In our opinion, it means only that these approaches to $1/m_Q$ corrections are doubtful, which is not surprising in view of the many approximations involved, in contrast with the rigorous treatment of the $m_Q \to \infty$ limit.

 In the case of the BBNS analysis, {as we have already noticed, }it is recognised by the authors themselves below their Eq. (28), as concerns the spectator contribution : it is noted there that the $\alpha_s(k^2)$ factor for the perturbatively treated gluon cannot be trusted since $k$ is small. The uncertainty cannot be estimated. On the other hand, as to Bordone et al. \cite{ref:facto1}, LCSR are commonly recognised to be rather uncertain.
This is implicitly acknowledged by the authors as they
include a factor $10$ of uncertainty.


\vskip 0.5cm


\subsection{Values for $| a_1 |$ with an emitted heavy meson}

{ We have found that factorisation still applies to a reasonable degree of approximation when the emitted meson is a charmed one. That is, $a_{1,eff}$ is found experimentally to be not too far from $1$, all the more when the penguin contributions are subtracted, as is necessary. We must stress that this is not what was expected from the theoretical statement in the heavy quark limit, since it applies only to light meson emission (see, in this respect,  section  3.5.1 of BBNS).  In the present case, one must emphasize that the main statement, namely that the factorisable contribution dominates in the heavy quark limit, is no longer guaranteed because this contribution is now suppressed by a power of $1/m_Q$  with respect to the estimate for light meson emission. Indeed, in the limit $m_Q \to \infty$ with $m_c/m_b$ fixed, one still gets  {$F_0 \to \mathrm{cste}$} at $q^2=m^2_D$, but $f_D \propto (m_c)^{-1/2}$ (if, on the other hand, we had kept $m_c$ fixed, we would have obtained  $F_0 \to (m_b)^{-3/2}$).
 }

\section{Conclusions}

{We have seen that,  while the $a_1$ evaluations obtained in the $m_Q \to \infty$ limit have reached the $1 \%$ accuracy, they differ from experimental values, obtained with a similar accuracy, by about $20\%.$  In fact, the progresses made  both in the experimental measurements and in the theoretical evaluation {\bf of the asymptotic limit} have created a gap between them which is so large that it implies the necessity of a safe evaluation of the $1/m_Q$ corrections. Let us stress indeed that this gap is observed in all considered channels, whether the  process is  purely of the  tree topology type or exchange terms are also present; this fact seemingly shows that the exchange contributions  are small. } But now, as we argued in the preceding section, such a safe evaluation is lacking: the two proposed evaluations, by BBNS \cite{ref:bbns} and by Bordone et al. \cite{ref:facto1} seem quite unsatisfactory.  \bf  Therefore also, one should not base ``Precision physics''  - for instance the precise evaluation of $f_s/f_d$ - on theoretical estimates of $\Bob \to D^{(*)+}P^-$ non leptonic decays.

}

{On the other hand, one can formulate some qualitative, hypothetical, statements 
on the basis of observation of the presently available values of  $a_1$.}

{1) The differences are similar whether a $D$ or a $D^*$ is produced, as expected in the $m_Q \to \infty$ limit.}

  {2) Similarly, some approximate 
$SU(3)$ symmetry properties appear.
One can reasonably suspect that $SU(3)$ breaking in the\
 ratios of various $a_1$ factors is moderate, and {smaller} than the departure from $1$ of the $a_1$'s themselves, but it is not a quantitative statement. }

It has been advocated that predicted ratios between $a_1$ values obtained in that limit can be more reliable than absolute $a_1$ values. As a result, some sort of universality of these corrections has been invoked to use them to relate $\Bdb \to D^{(*)+}K^-$ and $\Bsb \to D_s^{(*)+}\pi^-$ decay channels and to consider that $a_1(\Bdb,D^{(*)+}K^-)=a_1(\Bsb,D_s^{(*)+}\pi^-)$. This equality is used in one of the method to obtain $f_s/f_d$ at LHC. Meanwhile, at present, there is no theoretical proof, nor experimental control, that such an equality is valid. We advocate to use the direct measurement of $a_1(\Bsb,D_s^{(*)+}\pi^-)$ using exclusive nonleptonic and semileptonic decays to test this hypothesis.

{  Finally we want to stress  the  unexpected conclusion that charmed meson emission obeys also factorisation with a coefficient quite similar to class I light meson emission, i.e. not too far from $1$. This is a purely empirical fact, not expected from BBNS asymptotic theory, which suggests that factorisation should not extend to heavy meson emission. }

\end{document}